%Paper: cond-mat/9210015
%From: jbm@yollabolly.physics.brown.edu (Brad Marston)
%Date: Sat, 17 Oct 92 11:24:43 -0500

%%%%%%%%%%%%%%%%%%%%%%%%%%%%%%%%%%%%%%%%%%

% djnlx.tex

%%%%%%%%%%%%%%%%%%%%%%%%%%%%%%%%%%%%%%%%%%
%%                      	DJNLX.TEX
%%
%%      This is an extended version of JNL.TEX Version 0.3 (as of 6/12/85).
%%
%%	This is a set of TeX 82 macros designed to produce scientific
%%	papers with a minimum of fuss and using as much of plain.tex as
%%	possible.  The user need only know what is in the TeXbook, and
%%	the macros under ``user definitions'' below.  Also, the user
%%	definitions are intended to be as simple as possible, so that
%%	the user may change them as desired.
%%

%%
%%  Font definitions suitable for the IMAGEN (Written by Tony Kennedy).
%%
%%  Boldface mathematic italic (\mib) is included. Sans serif, boldface
%%  sans serif, sans serif italic, "Q"-sans serif, "Q"-sans serif italic,
%%  "MC"-sans serif, Dunhill style, and "capitals and small capitals" style
%%  fonts can be included from external files by a single control sequence.
%%  Have a look at the JNLX.DOC file for further information and at the
%%  JNLXAMP.DVI file for examples.
%%
%%  (by Ulrich Kettler 5/8/86)(last change 6/10/86).
%%  (eleven point font by Chris Stanton)

%  Define a whole menagerie of pseudo-12pt fonts

         \font\twelvei=cmmi10 scaled 1200
  \font\twelvesy=cmsy10 scaled 1200

  \font\twelvemib=cmmib10 scaled 1200
  \font\elevenmib=cmmib10 scaled 1095
  
  \font\eightmib=cmmib10 scaled 800
  \font\sixmib=cmmib10 scaled 667

%  Define a whole menagerie of pseudo-11pt fonts

\font\elevenrm=cmr10 scaled 1095    \font\eleveni=cmmi10 scaled 1095
\font\elevensy=cmsy10 scaled 1095   \font\elevenex=cmex10 scaled 1095
\font\elevenbf=cmbx10 scaled 1095   \font\elevensl=cmsl10 scaled 1095
\font\eleventt=cmtt10 scaled 1095   \font\elevenit=cmti10 scaled 1095

% other fonts

\font\seventeeni=cmmi10 scaled \magstep3

\font\seventeensy=cmsy10 scaled \magstep3

\font\seventeenmib=cmmib10 scaled \magstep3

\newfam\cpfam%

%  Define a whole menagerie of pseudo-8pt fonts

\font\eightrm=cmr8 \font\eighti=cmmi8
\font\eightsy=cmsy8 \font\eightbf=cmbx8

%  Define a whole menagerie of pseudo-6pt fonts

\font\sixrm=cmr6 \font\sixi=cmmi6
\font\sixsy=cmsy6 \font\sixbf=cmbx6

\skewchar\eleveni='177   \skewchar\elevensy='60
\skewchar\elevenmib='177  \skewchar\seventeensy='60
\skewchar\seventeenmib='177
\skewchar\seventeeni='177

\newfam\mibfam%

%  elevenpoint

\def\elevenpoint{\normalbaselineskip=12.2pt
  \abovedisplayskip 12.2pt plus 3pt minus 9pt
  \belowdisplayskip 12.2pt plus 3pt minus 9pt
  \abovedisplayshortskip 0pt plus 3pt
  \belowdisplayshortskip 7.1pt plus 3pt minus 4pt
  \smallskipamount=3.3pt plus1.1pt minus1.1pt
  \medskipamount=6.6pt plus2.2pt minus2.2pt
  \bigskipamount=13.3pt plus4.4pt minus4.4pt
  \def\rm{\fam0\elevenrm}          \def\it{\fam\itfam\elevenit}%
  \def\sl{\fam\slfam\elevensl}     \def\bf{\fam\bffam\elevenbf}%
  \def\mit{\fam 1}                 \def\cal{\fam 2}%
  \def\tt{\eleventt}
  \def\mib{\fam\mibfam\elevenmib}%
  \textfont0=\elevenrm   \scriptfont0=\eightrm   \scriptscriptfont0=\sixrm
  \textfont1=\eleveni    \scriptfont1=\eighti    \scriptscriptfont1=\sixi
  \textfont2=\elevensy   \scriptfont2=\eightsy   \scriptscriptfont2=\sixsy
  \textfont3=\elevenex   \scriptfont3=\elevenex  \scriptscriptfont3=\elevenex
  \textfont\itfam=\elevenit
  \textfont\slfam=\elevensl
  \textfont\bffam=\elevenbf \scriptfont\bffam=\eightbf
  \scriptscriptfont\bffam=\sixbf
  \textfont\mibfam=\elevenmib
  \scriptfont\mibfam=\eightmib
  \scriptscriptfont\mibfam=\sixmib
  \def\xrm{\textfont0=\elevenrm\scriptfont0=\eightrm
      \scriptscriptfont0=\sixrm}
  \normalbaselines\rm}

  \skewchar\twelvei='177   \skewchar\twelvesy='60
  \skewchar\twelvemib='177
%
%  twelvepoint
%

%	tenpoint

%%
%%	Change of the internal codes for lowercase Greek letters
%%	in order to use {\mib\alpha}... for boldface "alpha"... .
%%
\mathchardef\alpha="710B
\mathchardef\beta="710C
\mathchardef\gamma="710D
\mathchardef\delta="710E
\mathchardef\epsilon="710F
\mathchardef\zeta="7110
\mathchardef\eta="7111
\mathchardef\theta="7112
\mathchardef\kappa="7114
\mathchardef\lambda="7115
\mathchardef\mu="7116
\mathchardef\nu="7117
\mathchardef\xi="7118
\mathchardef\pi="7119
\mathchardef\rho="711A
\mathchardef\sigma="711B
\mathchardef\tau="711C
\mathchardef\phi="711E
\mathchardef\chi="711F
\mathchardef\psi="7120
\mathchardef\omega="7121
\mathchardef\varepsilon="7122
\mathchardef\vartheta="7123
\mathchardef\varrho="7125
\mathchardef\varphi="7127

%%	Change of the internal codes of the uppercase Greek letters, in order
%%	to get mathematic italic "Gamma", by typing "\Gamma", and
%%	to get roman "Gamma", by typing "{\rm\Gamma}".
%%

%%	Various internal macros
%%

\def\beginlinemode{\endmode
  \begingroup\parskip=0pt \obeylines\def\\{\par}\def\endmode{\par\endgroup}}
\def\beginparmode{\endmode
  \begingroup \def\endmode{\par\endgroup}}
\let\endmode=\par
{\obeylines\gdef\
{}}
\def\singlespace{\baselineskip=\normalbaselineskip}

\def\oneandahalfspace{\baselineskip=\normalbaselineskip
  \multiply\baselineskip by 3 \divide\baselineskip by 2}
\def\doublespace{\baselineskip=\normalbaselineskip \multiply\baselineskip by 2}

\nopagenumbers
\newcount\firstpageno
\firstpageno=2
%% FOLLOWING LINE CANNOT BE BROKEN BEFORE 80 CHAR
\footline={\ifnum\pageno<\firstpageno{\hfil}\else{\hfil\elevenrm\folio\hfil}\fi}
\let\rawfootnote=\footnote		% We must set the footnote style
\def\footnote#1#2{{\oneandahalfspace\parindent=0pt
\rawfootnote{#1}{#2}}}
\def\raggedcenter{\leftskip=4em plus 12em \rightskip=\leftskip
  \parindent=0pt \parfillskip=0pt \spaceskip=.3333em \xspaceskip=.5em
  \pretolerance=9999 \tolerance=9999
  \hyphenpenalty=9999 \exhyphenpenalty=9999 }
\def\dateline{\rightline{\ifcase\month\or
  January\or February\or March\or April\or May\or June\or
  July\or August\or September\or October\or November\or December\fi
  \space\number\year}}
\def\received{\vskip 3pt plus 0.2fill
 \centerline{\sl (Received\space\ifcase\month\or
  January\or February\or March\or April\or May\or June\or
  July\or August\or September\or October\or November\or December\fi
  \qquad, \number\year)}}

%%
%%	Page layout, margins, font and spacing (feel free to change)
%%

\hsize=6.5truein
\hoffset=0truein
\vsize=8.9truein
\voffset=0truein
\hfuzz=0.1pt
\vfuzz=0.1pt
\parskip=\medskipamount
\overfullrule=0pt	% delete the nasty little black boxes for overfull box

%%
%%	The user definitions for major parts of a paper (feel free to change)
%%

	% Preprint number at upper right of title page

\def\title			%  Title on title page
  {\null\vskip 3pt plus 0.2fill
   \beginlinemode \doublespace \raggedcenter \bf}

\def\author			%  Author(s) name(s)  on title page
  {\vskip 3pt plus 0.2fill \beginlinemode
   \singlespace \raggedcenter}

\def\affil			% Affiliations (can intermix with \author)
  {\vskip 3pt plus 0.1fill \beginlinemode
   \oneandahalfspace \raggedcenter \sl}

\def\abstract			% Begin abstract
  {\vskip 3pt plus 0.3fill \beginparmode
   \doublespace \narrower ABSTRACT: }

\def\summary			% same as abstract
  {\vskip 3pt plus 0.3fill \beginparmode
   \doublespace \narrower SUMMARY: }

\def\pacs#1
  {\vskip 3pt plus 0.2fill PACS: #1}

\def\endtitlepage		% End title page, begin body of paper
  {\endpage			% 	This subsumes \body
   \body}

\def\body			% Begin text body;  can be used to end
  {\beginparmode}		% \title, \author, \affil, \abstract,
				% \reference, or \figurecaption modes

\def\head#1{			% Head;  NOTE enclose the text in {}
  \filbreak\vskip 0.5truein	%  e.g., \head{I. Introduction}
  {\immediate\write16{#1}
   \raggedcenter \uppercase{#1}\par}
   \nobreak\vskip 0.25truein\nobreak}

\def\refto#1{$^{#1}$}		% For references in text as superscript

\def\references			% Begin references -- basic format is Phys Rev
  {\head{References}		% I.e., volume, page, year (space after commas).
   \beginparmode
   \frenchspacing \parindent=0pt \leftskip=1truecm
   \parskip=8pt plus 3pt \everypar{\hangindent=\parindent}}

\gdef\refis#1{\indent\hbox to 0pt{\hss[#1]~}}	% Ref list numbers.

\gdef\journal#1, #2, #3, 1#4#5#6{		% Journal reference.  Comma sets
    {\sl #1~}{\bf #2}, #3 (1#4#5#6)}		% off: name, vol, page, year

\def\refstylenp{		% Nucl Phys(or Phys Lett) ref style: V, Y, P
  \gdef\refto##1{ [##1]}				% Reference in text []
  \gdef\refis##1{\indent\hbox to 0pt{\hss##1)~}}	% Ref list numbers)
  \gdef\journal##1, ##2, ##3, ##4 {			% Journal reference
     {\sl ##1~}{\bf ##2~}(##3) ##4 }}

\def\refstyleprnp{		% Input like pr, output like np!!
  \gdef\refto##1{ [##1]}				% Reference in text []
  \gdef\refis##1{\indent\hbox to 0pt{\hss##1)~}}	% Ref list numbers)
  \gdef\journal##1, ##2, ##3, 1##4##5##6{		% Journal reference
    {\sl ##1~}{\bf ##2~}(1##4##5##6) ##3}}

\def\refstylejphys{		% Input like pr, output like np!!
  \gdef\refto##1{[##1]}				% Reference in text []
  \gdef\refis##1{\indent\hbox to 0pt{\hss[##1]~}}	% Ref list numbers)
  \gdef\journal##1, ##2, ##3, 1##4##5##6{		% Journal reference
     1##4##5##6 {\sl ##1~}{\bf ##2~}{##3}}}

\def\endreferences{\body}

\def\figurecaptions		% Begin figure captions
  {\endpage
   \beginparmode
   \head{Figure Captions}
}

\def\endpage			%  Eject a page
  {\vfill\eject}

\def\endpaper			%  Ways to say goodbye
  {\endmode\vfill\supereject}

%%
%%	Various little user definitions
%%

\def\ref#1{Ref[#1]}			% 	for inline references
\def\Ref#1{Ref[#1]}			% 	ditto
\def\Refs#1{Refs[#1]}			% 	ditto
			% 	ditto
			% 	ditto

		% For citation of equation numbers
	%	ditto
			%	ditto
			%	ditto
			%	ditto
			%	ditto
\def\frac#1#2{{\textstyle{#1 \over #2}}}
\def\half{{\textstyle{ 1\over 2}}}

\def\sla{\raise.15ex\hbox{$/$}\kern-.57em}
\def\leaderfill{\leaders\hbox to 1em{\hss.\hss}\hfill}
\def\twiddle{\lower.9ex\rlap{$\kern-.1em\scriptstyle\sim$}}
\def\bigtwiddle{\lower1.ex\rlap{$\sim$}}
\def\gtwid{\mathrel{\raise.3ex\hbox{$>$\kern-.75em\lower1ex\hbox{$\sim$}}}}
\def\ltwid{\mathrel{\raise.3ex\hbox{$<$\kern-.75em\lower1ex\hbox{$\sim$}}}}
\def\square{\kern1pt\vbox{\hrule height 1.2pt\hbox{\vrule width 1.2pt\hskip 3pt
   \vbox{\vskip 6pt}\hskip 3pt\vrule width 0.6pt}\hrule height 0.6pt}\kern1pt}

\catcode`@=11
\newcount\r@fcount \r@fcount=0
\newcount\r@fcurr
\immediate\newwrite\reffile
\newif\ifr@ffile\r@ffilefalse
\def\w@rnwrite#1{\ifr@ffile\immediate\write\reffile{#1}\fi\message{#1}}

\def\writer@f#1>>{}
\def\referencefile{%			  Stuff to write .REF file
  \r@ffiletrue\immediate\openout\reffile=\jobname.ref%
  \def\writer@f##1>>{\ifr@ffile\immediate\write\reffile%
    {\noexpand\refis{##1} = \csname r@fnum##1\endcsname = %
     \expandafter\expandafter\expandafter\strip@t\expandafter%
     \meaning\csname r@ftext\csname r@fnum##1\endcsname\endcsname}\fi}%
  \def\strip@t##1>>{}}

\def\citeall#1{\xdef#1##1{#1{\noexpand\cite{##1}}}}
\def\cite#1{\each@rg\citer@nge{#1}}	% Variable No. of args, separated by ","

\def\each@rg#1#2{{\let\thecsname=#1\expandafter\first@rg#2,\end,}}
\def\first@rg#1,{\thecsname{#1}\apply@rg}	% each@ag is a general purpose
\def\apply@rg#1,{\ifx\end#1\let\next=\relax%	  variable no. of arg. macro.
\else,\thecsname{#1}\let\next=\apply@rg\fi\next}% args separated by commas

\def\citer@nge#1{\citedor@nge#1-\end-}	% Check for M-N range (M and N numbers)
\def\citer@ngeat#1\end-{#1}
\def\citedor@nge#1-#2-{\ifx\end#2\r@featspace#1 % Single argument
  \else\citel@@p{#1}{#2}\citer@ngeat\fi}	% M-N range of arguments
\def\citel@@p#1#2{\ifnum#1>#2{\errmessage{Reference range #1-#2\space is bad.}
    \errhelp{If you cite a series of references by the notation M-N, then M and
    N must be integers, and N must be greater than or equal to M.}}\else%
 {\count0=#1\count1=#2\advance\count1
by1\relax\expandafter\r@fcite\the\count0,%
  \loop\advance\count0 by1\relax%	  Loop from M to N
    \ifnum\count0<\count1,\expandafter\r@fcite\the\count0,%
  \repeat}\fi}

\def\r@featspace#1#2 {\r@fcite#1#2,}	% Eat spaces at beginning or end of arg
\def\r@fcite#1,{\ifuncit@d{#1}		% Cite individual reference
    \expandafter\gdef\csname r@ftext\number\r@fcount\endcsname%
    {\message{Reference #1 to be supplied.}\writer@f#1>>#1 to be supplied.\par
     }\fi%
  \csname r@fnum#1\endcsname}

\def\ifuncit@d#1{\expandafter\ifx\csname r@fnum#1\endcsname\relax%
\global\advance\r@fcount by1%
\expandafter\xdef\csname r@fnum#1\endcsname{\number\r@fcount}}

\let\r@fis=\refis			% Save old \refis, redefine
\def\refis#1#2#3\par{\ifuncit@d{#1}%      Use two params #2 #3 to strip blank
    \w@rnwrite{Reference #1=\number\r@fcount\space is not cited up to now.}\fi%
  \expandafter\gdef\csname r@ftext\csname r@fnum#1\endcsname\endcsname%
  {\writer@f#1>>#2#3\par}}

\def\r@ferr{\endreferences\errmessage{I was expecting to see
\noexpand\endreferences before now;  I have inserted it here.}}
\let\r@ferences=\references
\def\references{\r@ferences\def\endmode{\r@ferr\par\endgroup}}

\let\endr@ferences=\endreferences
\def\endreferences{\r@fcurr=0%		  Save old \endreferences, redefine
  {\loop\ifnum\r@fcurr<\r@fcount%	  Loop over refnum and produce text
    \advance\r@fcurr by 1\relax\expandafter\r@fis\expandafter{\number\r@fcurr}%
    \csname r@ftext\number\r@fcurr\endcsname%
  \repeat}\gdef\r@ferr{}\endr@ferences}

% Save old \endpaper, redefine it to write parting message.

\let\r@fend=\endpaper\gdef\endpaper{\ifr@ffile
\immediate\write16{Cross References written on []\jobname.REF.}\fi\r@fend}

\catcode`@=12

\citeall\refto		% These macros will generate citations
\citeall\ref		%
\citeall\Ref		%
\citeall\Refs		%

%
%  ion.tex:  use plain TeX with the above macro files
%
\elevenpoint\doublespace
\voffset 0.3in
\fontdimen13\twelvesy=5pt
\fontdimen14\twelvesy=5pt
\fontdimen15\twelvesy=5pt
\fontdimen16\twelvesy=5pt
\fontdimen17\twelvesy=5pt

\def\pb{\phantom{\bullet}\kern-3.5pt}

\title{Many Body Theory of Charge Transfer
in Hyperthermal Atomic Scattering}
\bigskip
\author{J. B. Marston}
\affil{
Department of Physics
Box 1843
Brown University
Providence, RI 02912 U.S.A.}
\author{D. R. Andersson}
\author{E. R. Behringer}
\author{B. H. Cooper}
\author{C. A. DiRubio}
\author{G. A. Kimmel}
\author{and}
\author{C. Richardson}
\affil{
Laboratory of Atomic and Solid State Physics
Cornell University
Ithaca, NY 14853-2501 U.S.A.}

\vfill\eject

\abstract{We use the Newns-Anderson
Hamiltonian to describe many-body electronic processes that occur when
hyperthermal alkali atoms scatter off metallic surfaces.
Following Brako and Newns, we expand
the electronic many-body wavefunction in
the number of particle-hole pairs (we keep terms up to and including
a single particle-hole pair).  We extend their earlier work by including
level crossings, excited neutrals and negative ions.
The full set of equations of motion are integrated numerically,
without further approximations, to obtain
the many-body amplitudes as a function of time.
The velocity and work-function dependence of final state quantities
such as the distribution of ion charges and excited atomic occupancies
are compared with experiment.  In particular, experiments that
scatter alkali ions off clean
Cu(001) surfaces in the energy range 5 to 1600 eV
constrain the theory quantitatively.
The neutralization probability of Na$^+$ ions
shows a minimum at intermediate velocity in agreement with
the theory.  This behavior contrasts with that of K$^+$, which shows
virtually no neutralization, and with Li$^+$, which exhibits a monotonically
increasing neutral fraction with decreasing velocity.  Particle-hole
excitations are left behind in the metal during a fraction of
the collision events;
this dissipated energy is predicted to be quite small (on the order of
tenths of an electron volt).
Indeed, classical trajectory simulations of the surface dynamics
account well for the observed energy loss, and thus provide some
justification for our
truncation of the equations of motion at the single particle-hole pair
level.  Li$^+$ scattering experiments off low work-function surfaces
provide qualitative information on the importance of many-body effects.
At sufficiently low work function, the negative ions predicted to occur
are in fact observed.  Excited neutral Li atoms (observed via
the optical 2p $\rightarrow$ 2s transition) also emerge from the
collision. A peak in the calculated Li(2p) $\rightarrow$ Li(2s)
photon intensity
occurs at intermediate work function in accordance with measurements.}

\pacs{34.70.+e, 79.20.Rf, 79.80.+w, 71.10.+x}

\endtitlepage

\centerline{I. INTRODUCTION}
\smallskip
The single-particle picture of resonant charge transfer,
based on a time-dependent Newns-Anderson Hamiltonian,
successfully explains the observed work-function dependence
of the neutralization probability of positive hyperthermal alkali
ions that sputter\refto{Yu} or
scatter\refto{LG,Kimmel} off metallic surfaces.  (For a review,
see \Ref{BN3}.)  The key
simplifying feature of this approximation is the absence of multiple
atomic degrees of freedom:
the electrons are treated as spinless Fermions that either occupy or
do not occupy a single valence orbital of the alkali ion.  (The Pauli
exclusion principle guarantees that double occupancy cannot occur.)
Analytical solutions to the
single-particle problem can be obtained.\refto{BN1}

Yet questions remain.
When the atomic orbital is degenerate, the single-electron picture
breaks down.  For example, the valence s-orbital of a positive alkali ion
may be filled with either a spin-up or a spin-down electron.  The
degeneracy is a non-trivial complication, because
strong correlations must exist.  In the alkali case,
once a spin-up electron transfers to the ion, subsequent attempts to transfer
a spin-down electron are discouraged
by the strong intra-atomic Coulomb repulsion between the two valence electrons.
(The repulsion manifests itself in the fact that
the electron affinity energy of alkali atoms is much smaller than the
ionization energy.)
The complication is reminiscent of the Kondo problem of a magnetic ion
embedded in a metal where the spin residing on the impurity couples to the
conduction electrons via virtual processes which allow a second electron
to temporarily jump onto the ion at some large energy cost.  The Kondo
effect is a collective phenomenon characterized by strong many-body
correlations induced by the impurity spins.
Thus the fact that real electrons come in two forms (spin up or down)
means that the single-particle picture really does not describe even
the simplest problem of a single atomic orbital.
It is therefore interesting to inquire into why the single-particle
results fit the neutralization experiments so well.

Multiple atomic orbitals are another source of degeneracy and correlations.
For example, the affinity p-orbitals of a neutral oxygen atom
are degenerate, at least when
the atom is far from the metal surface.  When one of these orbitals
acquires an electron, further transfers (which would yield an O$^{--}$ ion)
are energetically disfavored.
Langreth and Nordlander recently emphasized\refto{LN} that the neglect of
such correlations
can lead to qualitatively incorrect results.  For example, the p$_x$ and
p$_y$ orbitals of an oxygen atom
couple only weakly to a metal surface with its normal in the
$\bf z$ direction.  Therefore the p$_z$ orbital should fill first
as the atom approaches the surface.  Once filled, additional
electrons will be locked
out of the p$_x$ and p$_y$ orbitals by the Coulomb repulsion.
As the atom departs from the surface,
there will be ample time, if the atom is not traveling too fast,
for the p$_z$ orbital to empty,
yielding a neutral oxygen as the final state.
Had the intra-atomic
Coulomb energy been ignored, the final state would have been
a negative ion, because the p$_x$ and p$_y$ orbitals would also fill
when the atom is close to the metal and then retain their
electrons as the atom moves away.

The problem resembles the much-discussed ``Coulomb Blockade''\refto{CB}
which encumbers electrons that hop on to a small conducting dot of
capacitance $C$.
In the present case,
the atom functions as a capacitor because extra energy is required
to add a second electron.

To treat these many-body
correlations, we resort to an approximate solution of the
Newns-Anderson problem.
We employ a systematic
$1/N$ expansion ($N$ is the spin degeneracy of the electrons and
equals two for the physical case of spin up and down species)
to study the dynamics
of charge transfer involving
multiple orbitals.  The expansion is equivalent to a variational
expansion of the many-body wavefunction in the number of particle-hole
pairs.  It was employed
with success in the Kondo problem\refto{Kondo1} (the perturbation series
converge
rapidly when N is large enough).
Indeed, the $1/N$ expansion behaves qualitatively the same as the
exact Bethe-ansatz solution.\refto{Kondo2}
Brako and Newns\refto{BN2}
applied it to the dynamical
charge transfer problem in 1985.  We go further by including level
crossings,
excited atomic states, and affinity levels in the calculation.  We find that
the results closely match those of the
single-particle picture over a
broad range of parameters.
Apparently, the single-particle picture works so well because the
incorporation of spin and higher energy atomic states has little
effect on the neutralization probability.
On the other hand, the production of negative ions and
excited neutrals becomes significant at low work functions.
For these cases, the more complete theory is essential for a proper
description of the observable physics.

The basic idea behind the $1/N$ expansion is as follows: when $N$ is large,
the amplitude for each of the $N$ types of electrons to transfer to or from the
atom must be scaled back so that the overall charge-transfer rate
for {\it any} of the N types of electrons stays reasonable.
In this limit, the rate of formation of
particle-hole pairs becomes smaller and smaller because these excitations
are produced by processes in which an electron of a given species performs
not one but two hops:
once from a filled state in
the metal to the atom and then another hop back to the metal into a
state above the Fermi level.
Particle-hole pair formation therefore
becomes negligible in the $N \rightarrow \infty$
limit, and the many-body equations are simple.  The advantages of this
systematic solution of the many-body problem are two-fold: First, it is
straightforward
to identify the correction terms that appear at each order in the 1/N
expansion.  Second, we can test whether the 1/N expansion breaks
down as N decreases from infinity down to the physical value of 2 (see below).

The present work is similar in some respects to
calculations by Sulston and
collaborators.\refto{SAD}  These earlier calculations, however, included
neither particle-hole excitations nor excited atomic states in the
variational wavefunction.  Later calculations by the same group incorporated
these states\refto{SAD2} but all of the calculations employed
a ``local time approximation'' of untested
accuracy to simplify the equations of motion.
This approximation alters the normalization of the many-body
wavefunction
which consequently has to be renormalized periodically during the integration
forward in time.
We avoid approximations
of this sort by directly integrating the full set of equations of motion.
We show below that particle-hole pairs play
a crucial role in erasing memory of the initial state of the incoming
atom; thus it is important to include them.  The inclusion of
particle-hole pairs also
enables us to estimate the amount of energy dissipated by their formation.
It is therefore possible to
check, both theoretically and experimentally, the size of the errors
attending the 1/N expansion since the single particle-hole channels represent
corrections to the lowest-order ($N \rightarrow \infty$) solution.
Finally, by adding excited atomic states, we are able to make
additional contact with experiment (which can detect optical
transitions as the excited states decay).
Competition between negative and excited neutral final states
is important and explains newly obtained experimental data.

Many features of the theory can be tested experimentally.  The most
important unknowns are the set of distance-dependent couplings
between the atomic states and the metal.  We use recent first-principles
calculations (in a single-particle approximation) of the
couplings\refto{NT,PN}.
Scattering experiments off clean surfaces, by avoiding
the complicated local variations in the electrostatic potential
produced by adsorbates, yield quantitative information that
check the validity of these parameters.
Indeed, different alkali species (Li, Na, and K)
exhibit qualitatively different behavior and the theory must account for
these differences.  Also, measurements of negative ion
fractions and excited neutral yields (which become significant
at relatively high velocities and low work functions) directly test the
many-body features of the theory.  Finally, experiments that measure
energy dissipation during the collision process, combined with
classical trajectory simulations, provide upper bounds on
the amount of particle-hole excitations left behind in the metal.
These bounds can be compared to the predicted losses due to
electronic mechanisms.
No one experiment is sufficient to establish the credibility of a theory
with several parameters; rather a combination of tests is required.  We
make preliminary comparisons between our theory and several
experiments below.

In section (II) we discuss a generalized Newns-Anderson Hamiltonian for
resonant charge-transfer that includes level crossings, electron spin,
excited neutrals, and negative ion states.  The model serves as a
starting point for extensions to more complicated situations that will
be the focus of future work.  The systematic solution of
the many-body dynamics is presented in section (III).  Section (IV)
is devoted to a preliminary experimental evaluation of the theory.  We address
neutralization rates, the formation of
excited neutrals and negative alkali ions off low work function surfaces,
and the energy loss due to the formation of particle-hole pairs.
Conclusions and a discussion of open questions are presented in section (V).

\vfill\eject

\centerline{II. THE GENERALIZED NEWNS-ANDERSON MODEL}
\smallskip

To begin, we make several simplifying assumptions.  We employ the
Newns-Anderson Hamiltonian, ignore radiative and Auger
charge transfer processes, and consider only resonant charge transfer.
Charge transfer that involves the emission
of a photon is suppressed relative to resonant charge transfer
by a factor of $\alpha \approx 1/137$
(the fine-structure constant)
and the inclusion of Auger processes is something we plan to address in
future work.
The electrons in the target metal are modeled as
non-interacting spinning Fermions, albeit with renormalized parameters such as
effective mass.  Finally, the atom is modeled as a system with
a finite number of discrete states moving along a fixed classical trajectory
given by $z(t)$ where $z$ is the distance from the atom to the metal surface.
(For a jellium model of the metal electrons, $z$ is the distance
from the nucleus of the hyperthermal atom to the jellium edge.)
Each of these states couples to the metal electrons when the atom is
sufficiently close to the metal surface.
Feedback between the electronic degrees of freedom and the trajectory is
ignored in the formulation.  This approximation should be adequate as
long as the kinetic energy of the ion is much larger than the electronic
energies.

The model is defined by the following generalized time-dependent
Newns-Anderson Hamiltonian:
$$\eqalign{H(t) &= \sum_a [\epsilon^{(1)}_a(z) P_1  + \epsilon^{(2)}_a(z) P_2]
\ c_a^{\dagger \sigma} c_{a \sigma}
+ \sum_k \epsilon_k\ c_k^{\dagger \sigma} c_{k \sigma}\cr
&+ N^{-1/2} \sum_{a;\ k} \{ [V^{(1)}_{a;k}(z) P_1 + V^{(2)}_{a;k}(z) P_2]
\ c_a^{\dagger \sigma} c_{k \sigma} + H.c \} \cr
&+ \sum_{a > b} U_{ab} n_a n_b\ + \half~ \sum_a U_{aa} n_a (n_a - 1)\ .\cr}
\eqno(2.1)$$
Here $c_a^{\dagger \sigma}$ creates a spin $\sigma$ electron in
orbital $a$ of the atom (ie. for Li, a = 0 for the 2s orbital, a = 1, 2,
and 3 for
2p$_x$, 2p$_y$, and 2p$_z$, etc.)  Likewise, $c_k^{\dagger \sigma}$ creates an
electron of momentum $k$ and energy $\epsilon_k$
in the metal.  Of course, $k$ is really a three-vector,
but it may be regarded as a scalar without loss of generality by absorbing
the three-dimensional aspects of the problem into $\epsilon_k$
and $V_{a;k}$.  We introduce the operators $P_1$ and $P_2$ to
project respectively onto
atoms with one or two valence electrons.  These projectors, which may
be written in terms of the orbital occupancies
$n_a \equiv c_a^{\dagger \sigma} c_{a \sigma}$,
permit one to
assign different orbital energies ($\epsilon^{(1)}_a$ and $\epsilon^{(2)}_a$)
and metal-atom couplings
($V^{(1)}_{a;k}$ and $V^{(2)}_{a;k}$)
to the two cases of neutral atoms and negative ions.
An implicit sum over repeated upper and
lower Greek indices is adopted; for now $N = 2$ and $\sigma = 1,\ 2$
to represent the physical SU(2) case of spin up and down electrons.
Actually, when $N > 2$ additional projectors
$P_3, P_4,$ etc. should be included
to account for the possibility of having, say, three
SU(4) Fermions in the same orbital.  Instead, we implicitly assume
that these states have infinite energy and simply remove them from
the Hilbert space.  The removal of course has no effect on the physical
SU(2) results, but is just a formal trick to keep the $1/N$
expansion as simple as possible.  For convenience,  we also
multiply the atom-metal coupling by a factor of $N^{-1/2}$.  This factor
allows one to take the $N \rightarrow \infty$ limit without rescaling
$V_{a;k}$.
Finally, we neglect the possibility of spin-flip processes in our
Hamiltonian: $H$ is invariant under global SU(2) [or more generally
SU(N)] spin rotations.

$U_{ab}$ is the Coulomb repulsion between two
electrons in valence shells $a$ and $b$
which in principle depends on $z$ but which in practice we
assume to be constant.  (The assumption can be relaxed if necessary.)
As it stands, excited states
of negative ions are permitted.  But, because
these high energy states are not expected to play a significant role
in the many-body wavefunction, we eliminate them by taking
$U_{ab} \rightarrow \infty$ when orbitals $a$ and $b$ are not the lowest
s-orbital of the alkali atom.

We retain non-trivial time dependence in the orbital energies and
atom-metal couplings of the model.
The time-dependence enters through the ion trajectory, which we
sometimes model as:
$$\eqalign{z(t) &= z_f - u_i * t;\ t \leq t_{turn} \equiv (z_f - z_0)/u_i\ .\cr
&= z_0 + u_f*(t - t_{turn});\ t > t_{turn}.}\eqno(2.2)$$
Thus the trajectory starts at a distance $z_f$ far from the surface at
time $t=0$.  We account roughly for a decrease in the ion kinetic energy
(due principally to the recoil of surface atoms during impact) and the
change in the scattering angle here by instantaneously changing the initial
perpendicular
component of the ion velocity, $u_i$, to $u_f$ at the point of closest
approach,
$z_0$.
(Another possible trajectory, discussed below, neglects
the inward bound portion of the trip and instead starts the atomic motion
headed in an outward direction starting from the point of closest approach.)
More complicated time-dependent trajectories
can be incorporated as needed.  Attention must also be paid
to the dependence of the atom-metal
couplings and the effective density of states on
the {\it parallel} component of the ion velocity. Note that
the electronic states in the metal are shifted in momentum in the
reference frame of the ion\refto{WZ}.  For now we ignore parallel velocity;
the inclusion of this effect will be part of future refinements.

We define the Fermi energy to be zero and relate all other energies
to it; the vacuum level lies above it at energy $\Phi$, the work function.
Because of the image potential,
the ionization levels of the atom $\epsilon^{(1)}_a$ shift upward as it
approaches the metal surface:
$$\epsilon^{(1)}_a(z) = \epsilon_a(\infty) + \Phi + e^2/4z\ .\eqno(2.3)$$
Here $\epsilon_a(\infty)$ is the ionization energy of orbital $a$ of an
isolated atom, which is taken to be a negative number.
A more realistic model has the image shift saturate when the atom gets close
to the surface.
We account for the saturation by introducing a cutoff, $v_{max}$, in the
image potential.
Also, the image plane does not coincide exactly with the metal edge;
rather it can lie within a small distance of it.  Therefore we introduce
an adjustable parameter, $z_{im}$, the distance from the surface at which the
image saturates to value $v_{max}$.
So a better form for the ionization energy is given by:
$$\eqalign{\epsilon^{(1)}_a(z) &=
\epsilon_a(\infty) + \Phi +
(1/v_{max}^2 + 16 (z - z_{im})^2 / e^4)^{-1/2}\ ;\ z > z_{im}\cr
&= \epsilon_a(\infty) + \Phi + v_{max}\ ;\ z < z_{im}\ .\cr}
\eqno(2.4)$$
The two parameters in Eq. [2.4] can to some extent be determined
experimentally from an
analysis of the ion trajectories and energies\refto{Chris}.  We typically
take $v_{max} = 2.6$ eV and $z_{im} = 0.0$\AA\  for the Cu(001) surface.
Especially
interesting situations arise when the shift is large enough to push the
ground state ionization energy above the Fermi energy at
some crossing distance $z_c > z_0$.
In these cases, neutralization probabilities
can increase from nearly $0\%$ to $100\%$ as the velocity of outgoing
positive alkali ion decreases.\refto{Yu,Kimmel}
Local adsorbate induced electrostatic potentials are
obviously not included in Eq. [2.4].  Since adding adsorbates to the surface
is a convenient way of changing the work function,
it is often necessary to consider
local variations in the potential when fitting experimental results to
theory\refto{Geerlings,Kimmel}.
We propose to compute averages over different trajectories as part of
future work.

In contrast to the ionization levels, the affinity levels shift {\it downward}
by $e^2/4z$ as
the atom approaches the surface.  In other words, the energy required to
remove the two valence electrons bound to a negative ion
(thereby making it a positive ion)
is unaffected by the image charges.  Thus, it is simply:
$$\epsilon^{(2)}_a(z) = \epsilon_a(\infty) + \Phi\ . \eqno(2.5)$$
The intra-atomic Coulomb repulsion between two electrons in the
lowest s-orbital ($a = 0$) is then given by
$U = A - \epsilon^{(2)}_0(\infty) = A - \epsilon_0(\infty)$
where $U \equiv U_{00}$ and
$A$ is the electron affinity (also defined here to be negative).

The atom-metal couplings decay exponentially with distance when the atom
is far from the metal surface because the atomic wavefunctions
drop off exponentially with increasing distance from the atom
and the electronic wavefunctions in the metal fall off exponentially
with increasing $z$.
Closer in, the couplings deviate from the pure exponential form.
A systematic Laurent
expansion of the logarithm of the couplings (we suppress the occupancy
superscript) yields:
$$V_{a;k}(z) = \tilde{V}_{a;k}\ {\rm exp\ } [a_{-1}(a;k) / z + a_1(a;k) z]\ .
\eqno(2.6)$$  [Note that $V_{a;k}(z)$ need not be purely real.
Nevertheless, we take it to be real in the following calculations.]
Non-zero (and negative) $a_{-1}$ incorporates saturation in the
growth of the coupling at short distances.
Further terms $a_{-2},\ a_{-3},\ $ etc.
may be added to the Laurent expansion as needed.  In the
following calculations we generally ignore the $k$ dependence of
the metal-atom coupling.  This approximation is really quite severe.
It is justified in so far as most of the resonant electronic
processes occur close to the Fermi surface and the wavevector
dependence of the couplings is smooth.  Making this assumption
for the singly-occupied orbitals,
$$V^{(1)}_{a;k}(z) = \tilde{V}\
{\rm exp\ } [a_{-1}(a) / z + a_1(a) z]\ ,\eqno(2.7)$$
we find that
the functional form of Eq. [2.7] fits quite well values for $V^{(1)}(z)$
obtained from the single-particle widths calculated in \Ref{NT}
without recourse to additional terms in the Laurent expansion.

At far to moderate distances, we expect the overlap $V^{(2)}_0$
between the metal states and the affinity
orbital to be considerably larger than the overlap between the metal
and the neutral ground state orbital
because negative ions are very large in size.
Previous models (for example, those of \Ref{SAD2}, \Ref{Sebastian} and
\Ref{Kasai}) did not
account for this difference: the same
couplings were used for the affinity and ground state orbitals.
In fact the rms radius
of the Li$^-$ ion, calculated in a Monte-Carlo approach\refto{Cyrus}, exceeds
$2.0$\AA\ .  First-principle calculations of these couplings
are of course
desirable and we make use of recent computations by Nordlander\refto{PN}
(in a single-particle approximation) of the coupling between the metal
and negative ions.  Whether the single-particle approximation itself is
adequate for the calculation of these couplings is a question that requires
further investigation.

\vfill\eject

\centerline{III. A SYSTEMATIC SOLUTION}
\smallskip

Before embarking on the systematic solution to the many-body
Newns-Anderson system, we make some observations about other
theoretical approaches.
First it is clear that simply ignoring the intra-atomic Coulomb energy
in the physical problem with electrons of up and down spins
would give completely incorrect answers, even for the case of a single
atomic orbital.  For example, when a slow alkali ion bounces off a surface
with work-function $\Phi$ which is less than the magnitude of
the ionization energy, it should emerge neutralized:
an electron will always be able to transfer from the metal to the
valence orbital.  Under these conditions, neglecting $U$ would mean that
the atom actually emerges as a negative ion because if a spin-down
electron hops from the metal to the atom, so will a spin-up electron,
filling the orbital.  One might think that the intra-atomic repulsion
could be treated adequately in the
Hartree-Fock approximation.  But here again it is
impossible to get neutral fractions greater than $50\%$ because the two
spin species remain uncorrelated\refto{Sebastian}.  In other words, when
the neutral fraction becomes significant, so will negative ion formation.
This situation is at odds with
experiments that find nearly $100\%$ neutral fractions.

Exact diagonalization
of a Newns-Anderson Hamiltonian for targets consisting of
just three atoms in a chain is relatively straightforward\refto{Goldberg}.
But because the
Hilbert space becomes unmanageably large for more than a few atoms, and
because the existence of a continuum of metal states is required for an
adequate description of resonant charge-transfer, this
method cannot be applied to the macroscopic metal targets that are of
interest here.
Nevertheless, it was found that the single-particle approximation
describes the three-atom cluster reasonably well
when the ionization and affinity energies of the atoms are very
different\refto{Goldberg}.
This result anticipates our observation that both the single-particle
and many-body pictures yield similar values for the alkali neutralization
probability when the affinity levels lie well above the Fermi energy.

Successes in understanding
the Kondo problem
(the static limit of
the Newns-Anderson Hamiltonian) suggest some different approaches.
The slave-Boson Green's function
method is a convenient technique for enforcing the
constraint of single orbital occupancy in the $U \rightarrow \infty$ limit, and
Langreth and Nordlander apply it to the resonant charge-transfer
problem\refto{LN}.  In the limit of low ion velocity and high temperature
[$\Gamma(z) \beta << 2 \pi$ and $\alpha u << 2 \pi$ where $\beta$ is the
inverse temperature, $u$ is the velocity in atomic units, and the width of the
atomic levels is assumed to drop off exponentially:
$\Gamma(z) \propto {\rm exp}(-\alpha z)$]
they obtain simple coupled master equations from the low-order
equations for the occupancies of the atomic
orbitals.  The problem of finite intra-atomic Coulomb interactions may also
be treated by extensions of this approximate method\refto{Shao}.
Unfortunately,
the master equations are not justified at higher velocities; instead
cumbersome Dyson equations must be solved.

Since we are primarily interested in the case of higher
ion velocities (which enhance non-adiabatic survival
of excited neutrals and negative ions)
we prefer to follow the different, but related, systematic approach
of Brako and Newns\refto{BN2} and group
the full many-body electronic
wavefunction into sectors containing more and more numbers of
particle-hole excitations in the metal.  Upon truncating the wavefunction
at a given number of particle-hole pairs, we obtain a variational
wavefunction that spans only a tiny portion of the entire Hilbert space.
However, as long as the amplitude for the formation of particle-hole pairs
during the ion-surface collision
remains relatively small, we may view the wavefunction as a good
approximation to the full one.
(The expansion bears some resemblance to the ``equations of motion method''
employed by Kasai and Okiji\refto{Kasai} and the coupled-cluster expansion
of Sebastian\refto{Sebastian}.)
The amplitude for particle-hole pair
production
may be controlled at least formally by generalizing the
two types of SU(2) electrons (spin up and down) to N types of SU(N)
Fermions.  Thus the spin index $\sigma$ now runs from 1 to N.
We show below that the amplitudes for terms involving more and
more particle-hole pairs are reduced by higher and higher
powers of 1/N.  As long as
N is large enough, the errors introduced by the truncation of the
Hilbert space should be small.
We present theoretical and experimental
evidence to show that even in the physical case
$N = 2$ higher-order terms in the expansion are small.

To begin, we decompose the many-body wavefunction into four sectors
plus the remaining Hilbert space:
$$\eqalign{| \Psi(t) \rangle &= f(t) | 0 \rangle +
\sum_{a;\ k<k_f} b_{a;k}(t) |a; k \rangle
+ \sum_{k<k_f,\ l>k_f} e_{l,k}(t) |l, k \rangle
+ \sum_{q<k<k_f} d_{k,q}(t) |k, q \rangle\cr
&+ (rest\ of\ Hilbert\ space)\ .} \eqno(3.1)$$
Each sector is a global SU(N) singlet.  Non-singlet sectors can be ignored
in so far as the initial state of the system
(a closed shell positive alkali ion far away from an unperturbed, non-magnetic
metal) and the Hamiltonian are both SU(N) singlets.
Here the orthonormal
basis states in different sectors of the Hilbert space are given by:
$$\eqalign{
|a; k \rangle &\equiv N^{-1/2}\ c_a^{\dagger
\sigma} c_{k \sigma} | 0 \rangle\ .\cr
|l, k \rangle &\equiv N^{-1/2}\ c_l^{\dagger \sigma} c_{k \sigma} | 0 \rangle
\ .\cr
|k, q \rangle
&\equiv [N(N-1)]^{-1/2}\ c_{0}^{\dagger \alpha}
c_{k \alpha}
c_{0}^{\dagger \beta}
c_{q \beta} | 0 \rangle\ .} \eqno(3.2)$$
The reference state $| 0 \rangle$ represents a positive alkali ion
(ie. an empty valence shell) along with the non-interacting
Fermi-liquid at zero-temperature in the absence of any particle-hole
excitations.
The limits on the momenta ranges appearing in Eq. [3.1]
are shorthand notation for $\epsilon_q < \epsilon_k
< \epsilon_f$ and $\epsilon_l > \epsilon_f$
where $\epsilon_f \equiv 0$ is the Fermi energy.
In other words, $k$ and $q$ label
hole momenta, and $l$ labels particle momentum, so while
$|l, q \rangle$ is a positive ion plus a particle-hole pair, the
state $|k, q \rangle$ instead represents a negative ion with
two holes in the metal.
A schematic of the different sectors of the Hilbert space
is presented in Figure [1].
Note that excited negative ions do not appear in Eq. [3.2].  These
states are
removed from the Hilbert space by hand since (as discussed
above) we set $U_{ab}
\rightarrow \infty$ for $a, b \neq 0$.  We show below that terms involving
two or more particle-hole pairs constitute higher-order corrections dropped
in the approximate solution.

The time-dependent coefficients appearing in the many-body wavefunction
Eq. [3.1] are amplitudes for the following states:

\indent{(1) $f(t)$ ---
A positive ion with no excitations in the metal.  Note that
$f(t = -\infty) = 1$ describes the initial state of
an experiment which directs incoming positive ions against the metal
target.}

\indent{(2) $b_{a;k}(t)$ --- A neutral atom with orbital $a$ occupied
and a hole left behind in the metal at momentum $k$.}

\indent{(3) $e_{l,q}(t)$ --- A positive ion and a single
particle-hole pair in the metal (the electron has momentum $l$
and the hole has momentum $q$).}

\indent{(4) $d_{k,q}(t)$ ---
A negative ion with a double-occupied s-orbital ($a = 0$)
and two holes in the metal at momenta $k$ and $q$.}

\noindent The restriction to this trial basis is achieved
by projecting the Schrodinger equation
$i {d\over{dt}} \Psi = H \Psi$ onto each sector of the Hilbert space
and we obtain the following equations of motion:
$$\eqalign{i {d\over{dt}}f &= \sum_{a;\ k<k_f} V^{(1)*}_{a;k}\ b_{a;k}\ .\cr
i {d\over{dt}}b_{a;k} &= (\epsilon^{(1)}_a -
\epsilon_k)\ b_{a;k} + V^{(1)}_{a;k}\ f
+ \delta_{a,0}\ (1 - 1/N)^{-1/2}\ \sum_{q<k_f}
V^{(2)*}_{0;q}\ [\theta(k-q)\ d_{k,q} + \theta(q-k)\ d_{q,k}] \cr
&+ N^{-1/2} \sum_{l>k_f} V^{(1)}_{a;l}\ e_{lk}\ .\cr
i {d\over{dt}}e_{l,k} &= (\epsilon_l - \epsilon_k)\ e_{l,k} +
N^{-1/2} \sum_a V^{(1)*}_{a;l}\ b_{a;k} .\cr
i {d\over{dt}}d_{k,q} &= (2 \epsilon^{(2)}_0 - \epsilon_k - \epsilon_q + U)
\ d_{k,q}
+ (1 - 1/N)^{-1/2}
\ (V^{(2)}_{0;q}\ b_{0;k} + V^{(2)}_{0;k}\ b_{0;q})\ .\cr} \eqno(3.3)$$
The step function $\theta(x) = 1$ when $x > 0$; otherwise it is zero.  Its
appearance here is in keeping with the convention of dropping amplitudes
$d_{k,q}$ when $k < q$ since they are redundant (ie. $d_{k,q} = d_{q,k}$).
The logic behind the truncation scheme
becomes clear upon considering the nature of the off-diagonal
coupling [terms in the Hamiltonian proportional to $N^{-1/2}$].  These terms
couple adjacent sectors of the Hilbert space.  (By adjacent we mean sectors
that differ by at most one elementary excitation in the band like a hole
or a particle-hole pair.)
In fact, repeated
applications of the atom-metal coupling to the reference state $|0\rangle$
generates all the sectors in the full singlet many-body wavefunction.
Now each time $V_{a;k}$ acts, it brings along a factor of $N^{-1/2}$.  Thus
amplitudes for sectors involving multiple particle-hole pairs are weakly
coupled to lower order terms when N is large.  In particular, from
Eq. [3.3] it is clear
that the {\it amplitude} for a single particle-hole pair is reduced by a factor
of $N^{-1/2}$ in comparison to the amplitudes for the sectors with
no particle-hole pairs ($f$, $b_{a;k}$
and $d_{k,q}$).  The {\it probability} for a particle-hole pair is
therefore reduced by a factor of $1/N$.  By keeping this next-order term
one gains insight into the size of the errors produced by the truncation of
the Hilbert space.  It is also possible to estimate the amount of
energy lost during the collision process from the formation of particle-hole
excitations.

Actually, two other single particle-hole sectors appears at
$O(N^{-1/2})$ in addition to the $|l, q \rangle$ particle-hole sector with
its unoccupied atomic orbital.
Amplitudes for a particle-hole pair along with
singly and doubly occupied atomic orbitals should also be included at this
order.
Since these new sectors involve amplitudes with respectively three and four
momenta indices (the additional indices label the extra holes left behind
in the metal when electrons transfer to the atomic orbitals),
the numerical task of integrating the equations of
motion becomes too taxing (see below)
and we drop these sectors from further consideration.  In any case,
the negative-ion plus particle-hole sector probably contributes little
weight because of its high energy.  The neglect of the
neutral plus particle-hole sector presumably introduces larger errors.
Nevertheless, the theory does describe experiments that measure
collision energy losses for outgoing {\it positive} ions since
dissipation occurs via the $|l, q \rangle$ positive ion, particle-hole
sector.  We take up this analysis in section (IV).

Curiously, upon taking the $N \rightarrow \infty$ limit and eliminating
the double-occupied and excited neutral subspaces (by assigning to these
sectors very large energies), we find that
the equations of motion resemble those of the
Heisenberg operators $\hat{c}_a(t)$ and $\hat{c}_k(t)$ in the Brako-Newns
single-particle picture\refto{BN1}
upon identifying $f \leftrightarrow \hat{c}_a$
and $b_{0,k} \leftrightarrow \hat{c}_k$.   Appearances
are deceiving in this case, however, for two reasons.
First, equations [3.3] give the time
evolution of amplitudes (ie. c-numbers), {\it not} operators\refto{BN3}.
The physical
meaning of this distinction is as follows: in the $N \rightarrow \infty$ limit
of the many-body problem
there can be no particle-hole excitations as these amplitudes
are suppressed by a factor of $N^{-1/2}$.  But in the single-particle
picture, any number of particle-hole excitations appear because the final
state of the system at $t \rightarrow +\infty$
is a Slater determinant built up with creation operators
that are themselves linear combinations of the creation
operators at the initial time:
$| \Psi(\infty) \rangle = \hat{c}_a^{\dagger}(\infty)
\prod_k \hat{c}_k^{\dagger}(\infty) | 0 \rangle$ where
$\hat{c}_a(t) = \hat{U}(t)\ \hat{c}_a(0)\ \hat{U}^\dagger(t)$,
$\hat{c}_k(t) = \hat{U}(t)\ \hat{c}_k(0)\ \hat{U}^\dagger(t)$, and:
$$\hat{U}(t) \equiv \hat{T} {\rm exp\ } \{i
\int_0^t d\tau \hat{H}(\tau)\}\ \eqno(3.4)$$
is the time-evolution operator.

The second difference between the single-particle picture and the
$N \rightarrow \infty$ limit of the many-body equations
concerns the sum over momentum in the first of
Equations [3.3]: the sum extends only
over the momentum ($k$) of states below the Fermi energy whereas
in the single-particle picture the operator that
destroys a filled atomic orbital ($\hat{c}_a$)
couples to states both above and below
the Fermi surface.  One effect of the restriction on $k$ becomes clear
upon comparing the final outcomes from different
initial conditions to test whether ``loss-of-memory'' occurs.
The ``loss-of-memory hypothesis''
states that the final state of the outgoing atom
should be independent of
its initial state if the atom
stays in the region of strong coupling
to the metal for enough time to erase any memory of the initial state.
However, in the $N \rightarrow \infty$ limit,
loss-of-memory no longer occurs if the initial incoming state is that
of a neutral atom: the electron on the atom cannot jump into a metal state
above the Fermi surface (since it is not coupled to those states) but can only
fill the single unoccupied state below the Fermi surface (which has vanishing
measure in the continuum limit of an infinite number of metal states.)  Thus,
the atom emerges from the collision in a purely neutral state.  In
contrast, an incoming positive ion {\it can}
neutralize because all the electrons
below the Fermi surface are available for charge transfer.  Upon turning on
the coupling to
the particle-hole pairs (by returning to the physical case of
$N \rightarrow 2$), the
incoming neutral atoms can now
ionize because the valence electron is allowed to transfer
into the unfilled levels above the Fermi surface.

This behavior illustrates the importance of electron-hole pairs
to the loss-of-memory process.  Since we truncate the expansion at the single
particle-hole level, perfect loss-of-memory does not occur: the final state
occupancies depend to some extent on the initial state.\refto{BN2}
If more pairs
could be included, the loss-of-memory would presumably improve.
In practice, we
find the discrepancy often to be small; on the other hand one may choose a
better initial condition that incorporates the physics of loss-of-memory by
starting the integration with the atom-metal system
in its ground state at the point of
closest approach to the metal (see below).  The initial condition is justified
both by experiments which show that loss-of-memory occurs and by the
single-particle picture where memory of the initial state rapidly dwindles
as time progresses along a given trajectory.
Of course, the equilibrium ground state
is an inappropriate starting point
if one wishes to study the amount of energy dissipated during
the collision process due to the formation of electron-hole pairs which arise
during both the inbound and outbound portions of the trajectory.  Nevertheless,
integrations that start from the equilibrium ground state appear satisfactory
for the purposes of making
detailed comparisons to experiments that measure the final occupancy
probabilities.

The equations of motion are solved numerically by using a finite number,
L, of discrete
momenta (typically L = 100 which means 100 states above and 100 states
below the Fermi surface).
Because amplitudes $e_{l,k}$ and $d_{q,k}$ have two momenta indices,
on the order of tens of thousands of coupled differential
equations must be integrated forward along the trajectory.
We employ a fourth-order Runge-Kutta algorithm with
adaptive time steps.  The numerical task is simplified by making a change
of variables to remove the diagonal terms from the equations of motion.
Let:
$$\eqalign{f(t) &= F(t)\ .\cr
b_{a;k}(t) &= B_{a;k}(t)\ {\rm exp}\{i[\epsilon_k t - \phi_a(t)]\} .\cr
e_{l,k}(t) &= E_{l,k}(t)\ {\rm exp}\{i(\epsilon_k - \epsilon_l)t\} .\cr
d_{q,k}(t) &= D_{q,k}(t)\ {\rm exp}\{i[(\epsilon_k + \epsilon_q - U) t
- 2\phi_0(t)]\} .\cr
} \eqno(3.5)$$
Here, $\phi_a(t) \equiv \int_0^t \epsilon_a(\tau)\ d\tau$ is the
time-evolution phase for the decoupled, but image shifted, atomic orbital.
(Recall that the time dependence of $\epsilon_a$ comes indirectly from
the time-dependent position $z(t)$.  For the simple trajectories of Eq.
[2.2] and the image shift of Eq. [2.4], the phase integral $\phi_a(t)$ may be
evaluated analytically.)
In the new basis we find:
$$\eqalign{i {d\over{dt}}F &= \sum_{a;\ k<k_f} V^{(1)*}_{a;k}
\ {\rm exp}\{i[\epsilon_k t - \phi_a(t)]\}\ B_{a;k}\ .\cr
i {d\over{dt}}B_{a;k} &= V^{(1)}_{a;k}\ {\rm exp}\{i[\phi_a(t) - \epsilon_k
t]\}
\ F \cr
&+ \delta_{a,0}\ (1 - 1/N)^{-1/2}\ \sum_{q<k_f} V^{(2)*}_{0;q}\ {\rm exp}
\{i[(\epsilon_q - U) t -
\phi_0(t)]\}\ [\theta(k-q)\ D_{k,q}
+ \theta(q-k)\ D_{q,k}] \cr
&+ N^{-1/2} \sum_{l>k_f} V^{(1)}_{a;l}\ {\rm exp}\{i[\phi_a(t) -
\epsilon_l t]\}\ E_{lk}\ .\cr
i {d\over{dt}}E_{l,k} &= N^{-1/2} \sum_a V^{(1)*}_{a;l}\ {\rm exp}
\{i[\epsilon_l t - \phi_a(t)]\}\ B_{a;k} .\cr
i {d\over{dt}}D_{k,q} &= (1 - 1/N)^{-1/2}
\ V^{(2)}_{0;q}\ {\rm exp}\{i[\phi_0(t) + (U
- \epsilon_q) t]\}\ B_{0;k} \cr
&+ (1 - 1/N)^{-1/2}
\ V^{(2)}_{0;k}\ {\rm exp}\{i[\phi_0(t) + (U - \epsilon_k) t]\}
\ B_{0;q}\ .\cr} \eqno(3.6)$$
[Actually, for discrete momenta, the amplitude $D_{k,k}$ is a special case
that must be
treated separately.  Factors of $\surd\overline{2}$ appear to keep the basis
given by Eq. [3.2] normal when $k = q$.  For simplicity, we
suppress these complicating factors here
(which can be neglected in the continuum limit
of an infinite number of momenta).]
Because the right hand sides of Eqs. [3.6] vanish
as the atom and the metal decouple, the equations of motion may
be integrated forward in time rapidly when the atom is far from the surface.
Probability must be conserved and we check that
$$1 = |F(t)|^2 + \sum_{a;\ k<k_f} |B_{a;k}(t)|^2 + \sum_{l>k_f,\ q<k_f}
|E_{l,q}(t)|^2 +
\sum_{q<k<k_f} |D_{k,q}(t)|^2 \eqno(3.7)$$ remains satisfied to within
desired numerical accuracy (typically better than 1 part in $10^5$) over
the entire course of the integration.

We choose either of
two initial conditions: (1) start the trajectory far away
from the metal surface (see Eq. [2.2]) or
(2) from the point of closest approach.
In case (1) the initial conditions are given by setting all the initial
amplitudes equal to zero with exceptions $F(t=0) = 1$ if the incoming
atom is a positive ion or $B_{0,0} = 1$ if it is neutral.  This starting
point is used in section (IV d) below
to evaluate the energy dissipation due to the formation of
particle-hole pairs during the collision.  In case (2)
the equilibrium ground state of the system is the starting
point.  (The ground state
is quickly obtained via the imaginary-time Lanczos algorithm.)  This
initial condition (which incorporates the physics of
complete loss-of-memory) appears to be best
for comparisons with experiments that make quantitative measurements of
charge fractions\refto{BN2} (see below).  The double-precision
computations are performed on IBM RS/6000 series machines\refto{JBM}.
Depending on the ion velocity, runs take from less than one
minute to over an hour of CPU time.  Errors introduced by approximating the
band continuum with a finite number of states are controlled in the
usual manner:  (1) runs with twice as many states must yield the
same final occupancies to within some tolerance and (2)
the final amplitudes $E_{l,k}$ and $B_{a;k}$
should be smooth functions of the momenta indices.  In particular, there
must be enough states near the Fermi surface to adequately sample the
various amplitudes.  We find that 100 states both above and below the band
are generally more than sufficient to sample the amplitudes.

The time-evolution of the occupancies
in the different many-body sectors for some typical runs
are presented in Figure [2].  (In this case, a lithium atom interacts with
a $r_s = 2.6$ jellium surface which has a work function of 4.0 eV.
The couplings between the atomic and metal states are given below
in section IV.)
The smoothness of the curves is one sign that
enough states have been included in the discrete metal band to adequately
emulate the continuum.
In Figure [2(a)] the incoming Li$^+$ ion
heads inwards towards the metal surface from a starting position
20 \AA\ from the surface and begins to
neutralize at around 6 \AA\ when the coupling to the metal becomes
sufficiently strong to permit an electron to transfer over to the ion.
At approximately 2.8 \AA\ the image shift is sufficiently large to push
the Li(2s) level above the Fermi energy.  At this point, electron probability
begins to transfer back to the metal, increasing the positive ion occupancy.
Particle-hole pair formation also becomes
appreciable because the electron on the atom can also transfer to a metal
level above the Fermi energy.  Only at the closest distances ($z < 2$ \AA\ )
does the
negative ion occupancy become appreciable.
At these distances the affinity level drops
below the Fermi level.  Apparently the negative sector competes with the
neutral and positive sectors at short distances, because the occupancy
in the neutral and positive sectors drops close in.
The excited neutral 2p$_z$ channel (not shown)
also becomes active at short distances.
On the outward leg of the journey, the positive channel
continues to grow until it reaches 2.8 \AA\ again and then electron probability
once again dumps back onto the atom, increasing the neutralization probability.
Finally, around 6 \AA\ the occupancies settle down.  We call this distance
the ``settling distance'' for the Li(2s) orbital.
This distance is to be distinguished from the ``freezing
distance'' which has been defined\refto{FD} to be the distance where the
charge state is determined.

In Figure [2a] the final probability for a
particle-hole pair to be formed during the interaction is about 4\%.
An average of 0.036 eV is dissipated due to these pairs.  Negative ion
and excited neutral
production at the relatively large work function of 4.0 eV is negligible;
these channels empty quickly as the atom
leaves the region of strong coupling.
As expected, particle-hole production is suppressed in Figure [2(b)]
(the particle-hole probability is about 0.2\% and
the average dissipated energy is only 0.002 eV)
since in this case the system starts from the equilibrium ground state at the
point of closest approach.  This low energy initial state is not
conducive to the formation of energetic particle-hole pairs.
It seems possible that the smaller occupancy in the particle-hole sector
for this initial condition increases the accuracy of our
particle-hole expansion and thus justifies our use of this initial condition
for comparisons with the charge state experiments.
Note that the two different initial conditions
yield similar final occupancies, to within about 13\%, for the positive and
neutral fractions, demonstrating that significant loss-of-memory occurs
despite the truncation of the Hilbert space at the one particle-hole
level.

The occupancy in the particle-hole channel for particle-hole pairs
of different energies
peaks near 0.6 eV for the run displayed in Figure [2(a)].
A peak in the particle-hole energy distribution is a generic feature
of our many-body theory; different parameters, however, change the value of
the peak energy. Similar peaks (and energy dissipations) were found
in the single-particle approximation of \Ref{Nak} and the calculation of
\Ref{SAD2}.

\vfill\eject

\centerline{IV. PRELIMINARY COMPARISON WITH EXPERIMENT}
\smallskip

In order to make contact with the experiments described below,
we make use of first-principles
calculations of the couplings between the atomic and metal states.
We first assume that
copper is adequately described by $r_s = 2.6$ jellium metal.
We then use level widths $\Delta_a(z)$ for neutral alkali atoms
calculated as in \Ref{NT}.  (Actually, the values reported
in \Ref{NT} are for $r_s = 2.0$.  The widths calculated for
$r_s = 2.6$ are very similar and these are the ones we use\refto{Nordlander}.)
For the negative ion width we use
values recently calculated for $r_s = 2.0$ by Nordlander\refto{PN}.
Similar values are also found via the coupled angular mode (CAM)
method\refto{Gauyacq}.
Next, we relate these level widths to the couplings $V_{a}(z)$ by the usual
single-particle
Fermi Golden-Rule formula: $\pi~ N~ L~ V_{a}^2 = \Delta_a~ D$ where D = 4.0 eV
is approximately the half-bandwidth of copper.  (This formula already
incorporates the factor of $N^\half$ that appears in the Hamiltonian
[2.1].)  By setting L = 100 and
fitting $V_a(z)$ to the form of Eq. [2.7] we obtain the following parameters
(all in atomic units).

\noindent (1) Lithium: coupling to the Li(2s) state.
$$\eqalign{\tilde{V} = {\rm exp}(-2.399)\ ,\cr
a_{-1} = -3.881\ ,\cr
a_1 = -0.4916\ .\cr}$$

Coupling to the Li(2p$_z$) state.
$$\eqalign{\tilde{V} = {\rm exp}(-4.183)\ ,\cr
a_{-1} = -0.7205\ ,\cr
a_{1} = -0.2346\ .\cr}$$

Coupling to the Li$^-$(2s$^2$) state.
$$\eqalign{\tilde{V} = {\rm exp}(-5.084)\ ,\cr
a_{-1} = +1.529\ ,\cr
a_{1} = -0.1669\ .\cr}$$

\noindent(2) Sodium: coupling to the Na(3s) state.
$$\eqalign{\tilde{V} = {\rm exp}(-2.121)\ ,\cr
a_{-1} = -5.557\ ,\cr
a_{1} = -0.4877\ .\cr}$$

Coupling to the Na(3p$_z$) state.
$$\eqalign{\tilde{V} = {\rm exp}(-3.062)\ ,\cr
a_{-1} = 0.531\ ,\cr
a_{1} = -0.1773\ .\cr}$$

Coupling to the Na$^-$(3s$^2$) state.
$$\eqalign{\tilde{V} = {\rm exp}(-5.151)\ ,\cr
a_{-1} = 1.677\ ,\cr
a_{1} = -0.1559\ .\cr}$$

\noindent The geometry and symmetry of the $p_x$ and $p_y$ orbitals
suggests that their coupling to the metal is small; this is borne out
by the jellium calculations\refto{NT}.
We ignore them in our analysis.

In the following four subsections we explore some consequences
of our many-body theory, keeping in mind the possibility that different
parameters could provide a better description of the observations.
(Nevertheless, these couplings serve as a standard of comparison for
future studies.)  We start our discussion with
a quantitative test of the theory: neutralization from a clean surface.
Local inhomogeneities in the surface potential are small for a clean surface,
making comparison with theory relatively easy.
We then consider two experiments
that directly test the many-body features of our theory: the detection of
excited neutrals and negative ions.  Finally, we discuss the formation of
particle-hole pairs.  In this case even qualitative comparisons are difficult;
we can only show that the predictions of the theory are consistent
with the experimental upper bound on energy dissipation.

\vfill\eject
\noindent{A. Neutralization by a Clean Copper Surface}
\smallskip

We measure neutralization probabilities
for Li, Na and K scattered from clean
Cu(001) along the $\langle 100 \rangle$ azimuth for a range of scattered
atom velocities.  Clean surfaces offer the advantage of minimizing
electrostatic inhomogeneities that complicate the interpretation of results.
Energetic considerations show that
the adiabatic charge states for Li and Na in the Li + Cu and Na + Cu systems
are neutral when the atoms are far from the surface, while for the K + Cu
system the K is positively ionized.
We find that the Li and K monotonically approach the adiabatic charge
states as the
scattered atom velocity decreases.
However, for Na the neutralization probability is
{\it nonmonotonic}; it initially decreases with decreasing velocity
and then increases, approaching the
adiabatic ground state only at the lowest velocities.

The experiments were performed in an ultra high vacuum (UHV) system.
The experimental techniques are
described else\-where.\refto{Cooper3,Cooper4,Cooper5}
Only the relevant details are
presented here.
In the experiments described in this section,
we used Li, Na, and K ions with incident energies
from 5 eV to 1600 eV.  All measurements were performed on a
clean Cu(001) surface, prepared by standard
sputter and anneal cycles.  The surface order and cleanliness were checked
with low energy electron diffraction (LEED)
and Auger electron spectroscopy (AES), respectively.  The
scattered atoms are detected with a time-of-flight (TOF) spectrometer,
mounted on a rotatable platform,
which can be used to make velocity- and angle-resolved measurements of
neutral and positively and negatively ionized alkalis.\refto{Cooper5}
This detector can be operated in a mode whereby the velocity- and
angle-resolved neutralization probabilities for the scattered alkalis can be
determined.

In Figure [3] we show the measured neutralization probability of
lithium and sodium scattering off of a clean
Cu(001)$\langle 100\rangle$ surface as a function of perpendicular
component of the outgoing atomic velocity.
The neutralization as a function of perpendicular velocity
$P^0(v)$ is
qualitatively different for each species.  For Li, the neutralization
monotonically decreases as the velocity increases and $0.25 \leq P^0
\leq 0.75$ for the velocities investigated.  (For K, essentially no
neutralization is found.)  For Na, the neutralization versus perpendicular
velocity has a minimum and $0.04 \leq P^0 \leq 0.15$ in the velocity
range investigated.

Figure [3] also shows predictions for the neutralization probability of lithium
and sodium
from both the many-body model and the standard single-particle
model\refto{Kimmel}
(with the same couplings as the many-body model, but
now of course only between the neutral s-orbital and
the metal).  Both theories reproduce the experimental trends.  The origin
of these trends becomes clear if we consider the different lifetimes and
energies of the Li(2s), Na(3s), and K(4s) states.  For potassium, the first
ionization energy is 4.34 eV.  Since the image potential increases the energy
of the K(4s) level, it lies predominantly above the Fermi level, and it
is energetically favorable for the K(4s) level to be empty over a wide
range of atom-surface separations.  Thus, almost no neutralization of K$^+$
will occur.  Indeed, the many-body theory also predicts little neutralization.
Lithium, on the other hand, has an ionization energy of 5.39 eV
and the neutralization probability decreases as the velocity increases.
The reason for this decrease is clear:
The Li(2s) level lies below the Fermi energy when the atom is far from the
surface; only close to the surface is it image shifted above the Fermi energy.
In the velocity range of the experiment,
the freezing distance decreases as the outgoing velocity increases,
enhancing the positive fraction.

Sodium is intermediate between these two cases.  The ionization energy
is 5.14 eV, so the Na(3s) resonance is, as in the case of lithium,
predominantly below the Fermi level far from the surface and predominantly
above it close to the surface.  However, because the Na(3s)
resonance is closer to the Fermi level there is considerably less
neutralization than for lithium and the neutralization does not monotonically
decrease as the scattered velocity increases.  The minimum observed in
Figure [3] is due to the approximately exponential increase in the level
width with decreasing atom-surface separation.  For sodium, this increase
is relatively more important at higher velocities than the shift in the
energy due to the image potential.  Thus, as the velocity increases and
the freezing distance decreases, even though the resonance is at a higher
energy, more of the resonance lies below the Fermi level and the
neutralization increases.

The curves for the many-body and single-particle models shown in Figure [3]
are qualitatively
similar.  However, for both Li and Na, the many-body model
predicts less neutralization at all velocities.  Four
possible reasons for the differences are given here.
First, differences between the two models indicate that inter-atomic
correlations are important, even in the dynamics of alkali scattering from
clean metal surfaces.  Consider the case of Li scattering
from clean Cu(001).  Different Li resonances, for example Li(2s) and Li(2p),
have different couplings to the surface and result in different
freezing distances along the outgoing trajectory.
The freezing distance for the
lowest energy state (the ground state neutral) is typically the smallest.
Thus, even at the distance where the occupation of the Li(2s) state is
frozen substantial occupation of the excited neutral state, Li(2p), can remain.
In fact, the many-body theory predicts that the occupation of the
Li(2p) state can be as high as $\sim$4\% at the settling distance of the
Li(2s) state.  As the atom moves further away from the surface, charge transfer
from the Li(2p) can now occur leaving the atom in a positively ionized state.
(Virtually no excited neutrals or negative ions survive far from the surface.)
Thus, if we ignore the role of spin, excited states and other channels
(single-particle picture) we will obtain incorrect occupation probabilities.

Second, the level widths used in the many-body model were
calculated in a single particle picture.\refto{NT,PN}  The use of these
widths in our many-body theory is not necessarily justified.  Indeed, if
we calculate the lifetimes of the various atomic states in our
many-body theory (by holding the atom at a fixed distance $z$ from the
surface), we obtain different widths than those obtained via
Fermi's Golden Rule which of course ignores correlations.  To see this, we
appeal to similarities in the $N \rightarrow \infty$ limit
between our many-body equations and
the equations for the time-evolution of the operators in the single-particle
picture mentioned above in section (III).  The equations are similar
only for atomic levels deep
below the Fermi energy.  Atomic widths for these levels
calculated in either picture are
the same in the $N \rightarrow \infty$ limit.  But upon taking the
physical $N \rightarrow 2$ limit we find additional terms (the particle-hole
amplitudes) arise in the many-body picture.  Thus atomic lifetimes calculated
in the two pictures will generally differ.  Perhaps a more sensible
approach would be to renormalize the metal-atom couplings $V_{a;k}$ to
make the many-body theory reproduce the lifetimes calculated
within the single-particle picture.  Investigations along these lines may
shed some light on the effect of many-body correlations on atomic lifetimes.

Third, the parallel velocity of the scattered atom
(which shifts the Fermi surface in the atomic reference
frame and changes the couplings)
must be incorporated into the model to obtain good
quantitative agreement between theory and data\refto{Cooper5}.
The parallel velocity effect is significant even at surprisingly
small velocities ($v \approx 0.01 v_f$) and non-glancing scattering
geometries (e.g.~$\theta_f = 45^\circ$).

Finally, our solution of the many-body model is approximate.  As mentioned
above, the final occupancies depend to some extent on the initial conditions.
This dependence on the initial conditions represents
a limitation of the approximate solution of the model since good
experimental and theoretical evidence exists for complete loss-of-memory.

Our comparison of the single-particle and many-body models
demonstrates that in the case of Li, Na and K
scattered from Cu(001) both theories agree
qualitatively with experiment.
Further comparison of experimental results with theoretical predictions
utilizing somewhat different parameters should provide additional
quantitative insight into the strengths of the couplings
between the metal and the atomic states.
Other experiments at lower work functions
highlight the differences between the single-particle
and many-body pictures. We describe two such experiments below.

\def\110{$\langle 1\bar 10\rangle$}
\def\001{$\langle 001\rangle$}
\def\az100{$\langle 100\rangle$}
\def\c001100{$\langle 100 \rangle$}
\def\cu001110{$\langle 110 \rangle$}

\def\today{\ifcase\month\or
  January\or February\or March\or April\or May\or June\or
  July\or August\or September\or October\or November\or December\fi
  \space\number\day, \number\year}

\vfill\eject
\noindent{B. Excited States}
\smallskip

We performed a number of experiments which directly test
the many-body aspects of our theory of resonant charge
transfer.  Using the TOF neutral spectrometer mentioned above and
low level photon counting techniques, we measured
the work function dependence
of both the relative yields of excited states and the charge state
fractions resulting from the scattering of low and hyperthermal
energy beams of alkali ions off an alkali-covered Cu(001) surface.
Production of these species is enhanced at low work functions and high
velocities (which shorten the freezing distances and thereby increase
the final occupancies of energetic states).
The theory predicts all of the qualitative trends exhibited by
the experimental data.

One feature of the theory presented here is that it
predicts the probability with which incident
ions are scattered into neutral excited states.  In this
section, we compare our theoretical predictions to measurements
of the relative Li(2p) yield produced when Li$^+$ strikes a
Cu(001) surface with sub-monolayer coverage of potassium adsorbates,
hereafter denoted as a
K/Cu(001) surface.  We measure the dependence of this yield on
the work function shift that is induced by depositing K
onto the Cu(001) surface.\refto{BACM}  Measurements
of this type have been made previously for Li$^+$ incident
on Cs/W(110).\refto{SHHMK}

To measure the relative yields of excited states, we collect
the photons which are emitted during the decay of these
states.  The photons are transported by a fiber optic cable
and counted by using a photomultiplier.  Line filters corresponding
to particular optical transitions can be inserted into the light path
to isolate the various excited states.

For the measurements presented here, the
energy of the impinging Li$^+$ ions is 400 eV and the incident angle
(measured with respect to the surface normal) is 65$^\circ$.
The incident beam is directed along the \az100 azimuth.  Ions
which are scattered into the Li(2p) state survive in the Li(2p)
state for a half-life of 27 nsec and decay to the
Li(2s) state by emitting a photon ($E=1.85$ eV, and
$\lambda = 673$ nm).  Thus, the ions scattered
into the Li(2p) state are detected by collecting the photons
corresponding to the Li(2p) $\rightarrow$ Li(2s) transition.

The single-particle picture of charge transfer,\refto{Kimmel}
shown schematically
in Figure [4] for the closed Li atom and clean Cu(001) system
with work function $\Phi = 4.59$ eV (this absolute value was
determined in \Ref{Work}) shows that
little scattering into the Li(2p) state and other,
higher-lying excited states is expected because
these states are not resonant with any occupied states in the metal.
However, as the work function decreases,
the occupied metallic states are brought into resonance with the
excited states.  Decreasing the work function therefore
increases the yield of excited atoms scattered from the surface.
Since the energy of the Li(2s) state is significantly
lower than that of the Li(2p) state, we expect the fraction
of atoms scattered into the Li(2s) state to be much larger than
the fraction of atoms scattered into the Li(2p) state.
In principle, excited states of higher energy may also
participate but should not constitute a
significant fraction of the excited states in the scattered flux.
(We have verified that greater than 90\% of the measured emitted light in the
optical range is due to the Li(2p) $\rightarrow$ Li(2s) transition.)
However, the
affinity level, also shown in Figure [4], will also be increasingly
populated as the work function decreases.
Competition between the Li(2p) and Li$^-$(2s$^2$) (i.e.,
negative ion) channels therefore should occur at low work functions.

In Figure [5], we plot the relative yield of Li$^+$ ions
which are scattered into the Li(2p) state versus the work
function shift induced by the deposition of K.  In the same figure,
we plot the predicted yield of Li(2p) at the maximum outgoing normal
velocity since (as explained above) these trajectories are
responsible for most of the excited states.  The theoretical results are
normalized to the experimental results (we comment on the absolute
numbers below).  Note that the peak values
of the measured and predicted yields occur at nearly the same value
of the work function shift, $\Delta\Phi \approx -1.8$ eV, and that
the peak in the measured yield is broader than that predicted
by the theory.

The peak in the predicted yield is due
to competition between the Li(2p) state and the Li$^-$(2s$^2$) state.
It seems that a balance between these
minority state populations obtains at work function values
near 2.6 eV.  This balance determines the work function value
at which the peak in the predicted Li(2p) yield occurs.
Our work indicates that this value is
relatively insensitive to the parameters we use in the many-body theory.
Other experiments on similar systems suggest that Auger processes may also
play a role.\refto{SHHMK}  We note that both mechanisms could be operating:
Auger neutralization may be occuring along the incoming portion of
the trajectory, but memory of the neutralization history will be erased
as the atom enters the strong coupling region.  On the outgoing trajectory,
the different resonant processes described by our theory
are consistent with experiments we have performed to date.  Future
extensions of the many-body theory that include Auger amplitudes
will address the question of the relative importance of Auger versus
resonant processes.

One likely explanation for the discrepancy between the widths of the
experimentally observed and theoretically predicted peaks is
our failure to account for local variations in the
electrostatic potential induced by the alkali adsorbates in the model.  Such
variations tend to smear out work-function dependence of
observable quantities like the neutralization
probability\refto{Geerlings,Kimmel}.  Also, the couplings between the
atomic states and the metal electrons may be altered significantly
in the vicinity of an adsorbate atom.  Local variations can be incorporated
into the model by averaging results over many possible ion trajectories
that impact the surface at different points and thus sample different
electrostatic potentials and couplings.

Finally, we estimate the fraction of atoms
scattered into the Li(2p) state to be of order 0.004 at
$\Delta\Phi \approx -1.8$ eV.  The peak occupancy of the 2p$_z$ state
predicted by the model is 0.026.
The predicted peak value is quite sensitive to the
atomic state lifetimes and energies near the surface.  Different
(but still reasonable)
values for these parameters change the excited state fraction by an
order of magnitude.

\vfill\eject
\noindent{C. Positive, Neutral, and Negative Ions}
\smallskip

To measure the charge state fractions
in the flux scattered into a particular final angle,
we use the TOF spectrometer.  Recall that
it permits discrimination and detection of
alkali particles with different charge states.
We measure the work function dependence of
charge state fractions in the scattered flux when Li$^+$ ions
impinge on Cs/Cu(001).\refto{BACM}
In our experiments, we direct a 400 eV Li$^+$ ion beam toward
the Cs/Cu(001) surface and along the \az100 azimuth, with an incident
angle of 65$^\circ$.
The final angle of the scattered particles was 64$^\circ$.  The
work function shift induced by the deposition of Cs on the
surface is measured with a Kelvin probe.
Similar measurements, but without velocity resolution, were reported
previously.\refto{Geerlings}

Figure [6] is a plot
of the measured charge state fractions versus the work function shift
induced by Cs adsorption.  When the surface is clean, the only
charge states in the scattered flux are the positive ion state
and the neutral
states; no negative ions are present
to within the experimental uncertainty of a few percent.
As the work function shift decreases from zero,
(in other words, as the work function decreases),
the positive ion fraction decreases;
a corresponding increase is seen in the neutral fraction.
For work function values greater than
approximately 2.6 eV, the negative ion
fraction is less than a few percent.  The negative ion fraction
becomes appreciable only for work function values less than 2.6 eV,
with a maximum value of 0.14.  In the range of work function
values for which the negative ion fraction is appreciable, the positive
ion fraction does not exceed a few percent.

Figure [6] shows that the qualitative trends displayed
by the charge state fractions are reproduced by the model.
We can qualitatively understand these trends within the one-electron picture.
First, consider the work function
dependence of the positive ion fraction.  As the work function decreases,
more and more of the atomic resonance corresponding to the
Li(2s) level lies below the Fermi level, leading to its increased
population (smaller positive ion fraction).
We can also construct a one-electron picture for the affinity
level\refto{GRLG2}.  As the work function decreases, an
increasing portion of the atomic resonance corresponding to
the Li$^-$(2s$^2$) level lies below the Fermi level, and it
should be increasingly populated and more negative
ions emerge from the collision.
Note that the slope of the measured ion fraction versus the work function
is smaller than that predicted by the model.
As in the previous section this is consistent with the
neglect of local variations in the electrostatic potential induced
by the adsorbates\refto{Kimmel,Geerlings}.

The above results are similar to those obtained by Geerlings {\it et al.}
for Li$^+$ scattering from Cs/W(110).  In experiments by Brenten {\it et al.}
for Li$^+$ incident on Cs/W(110), the relative yields of
Li$^+$ and Li$^-$ have been measured as a function of Cs coverage,
along with the yields of Li(2p) and emitted electrons.\refto{BMKKSK}
In addition, Ashwin and Woodruff have reported measurements
of the ratio of positive ion fractions for Li$^+$ scattering from Cu and Cs
when Li$^+$ is incident on Cs/Cu(110).\refto{Ashwin}

In summary, many final atomic states occur
when Li$^+$ scatters from a Cs-covered copper surfaces.
We observe Li$^+$, Li(2s), Li(2p), and Li$^-$(2s$^2$).
Auger processes also play a role in charge transfer
in these systems.\refto{BMKK}  We discuss this channel and its
incorporation into the many-body model (something not possible
in single-particle models) in the conclusion.

\vfill\eject
\noindent{D. Dissipation Due to the Formation of Particle Hole Pairs}
\smallskip

The importance of particle-hole pair formation in the scattering
of alkali ions from metal surfaces can be estimated by comparing the
measured final energies of ions scattered from a metal surface at incident
energies ranging from a few eV to a few keV to the final
energies predicted by classical trajectory simulations.  We assume here
that resonant charge transfer is the only significant mechanism for the
production of particle-hole pairs.  If, as the 1/N particle-hole
expansion assumes, particle-hole pair production is limited then trajectory
simulations which do not include energy loss to particle-hole pair
production should be able to reproduce the measured energy loss
in ion-surface collisions. In this section we describe experiments (for
more details see \Ref{Chris})
which make this comparison and which
lend credence to the assumption that particle-hole pair production is
limited.

We measured a series of energy spectra for Na$^+$
scattering from Cu(001) along the \az100 azimuth at an incident angle
of 45 degrees and a final angle of 45 degrees as measured from the surface
normal. The scattering takes place within the plane defined by the \az100
azimuth and the surface normal.
Trajectory simulations
indicate that these energy spectra contain contributions from four different
kinds of scattering trajectories. The highest energy peak contains
contributions from two different types of trajectories that coincidentally
scatter with nearly the same final energy, namely the quasi-double (QD) and
triple zig-zag (TZZ) trajectories. The middle and lowest energy peaks
consist of double zig-zag (DZZ) and quasi-single (QS) trajectories,
respectively.  The QS and QD trajectories involve momentum transfer to
atoms which lie along a \az100 chain.  The QS trajectory transfers
momentum to primarily one surface atom and the QD trajectory transfers
momentum to two adjacent surface atoms in the chain.  The zig-zag
trajectories scatter from atoms in adjacent \az100 chains. The DZZ
and TZZ trajectories involve respectively two or three surface atoms.

The relative cross sections (peak heights) and fractional energy
losses (peak energies) in the measured energy spectra change as a
function of incident energy. Both of these trends are reproduced
quantitatively  using the classical trajectory simulation code
SAFARI\refto{safari}.
SAFARI integrates Hamilton's equations of motion for the ion interacting
with the surface. The ion-surface interaction potential is one of the
input parameters in the simulation. The energy and angular distributions
of 10 to 100 eV Na$^+$ scattering from Cu(001) \az100 are reproduced
quantitatively using an interaction potential which consists of a sum of two
terms. The first contribution is a repulsive potential that
is modeled as a sum of
Hartree-Fock pair potentials where the sum runs over six or more
surface atoms closest to the scattering ion. To this repulsive term
a second, attractive, potential is added (see Eq. [2.4])
to account for the image interaction between
the ion and surface. The only free parameters in the potential are $V_{max}$
and the position of the image plane. The values of these parameters
($V_{max} = 2.6$ eV and the image plane is set at a distance
$0.8$ \AA\ beyond the first plane of copper nuclei)
are determined by requiring that the simulated energy and
angular distributions agree with the data.

In Figure [7] the ratio of the scattered energy to the incident energy,
$E/E_0$, of the peaks in the measured energy
spectra are plotted as a
function of the incident energy $E_0$.  Also shown are the values of the
corresponding
scattered trajectories calculated in the simulations
(solid triangles and line).
The measured energies are uncertain to within $\pm 0.5$ eV due to contact
potential differences within the apparatus.
The only energy loss mechanism included in these
simulations is momentum transfer from the scattering ion to the recoiling
surface atoms.  It must be noted that we assume that the
ion-surface potential is accurately modeled in our
simulation\refto{Chris}, since an increase in the depth of the attractive well
could mimic energy loss due to particle-hole formation.
With this in mind, the excellent agreement between the measured
and simulated energy loss of these trajectories suggests that the additional
energy transfer from the ion to the surface due to particle-hole pair
excitations in the metal is very small. This is consistent with the
theoretical conclusion that the energy dissipated due to the formation
of particle-hole pairs is limited to less than a few tenths of an
electron volt.  Thus we have some direct experimental evidence
that the systematic expansion in the number of particle-hole pairs
is well behaved.

\vfill\eject

\centerline{V. CONCLUSION}
\smallskip
The preceding sections describe a generalized Newns-Anderson
model and its systematic solution.  The theory goes beyond earlier work
in that it incorporates electron spin, Coulomb repulsion, level
crossings, particle-hole pairs, and excited atomic states all within one
systematic framework.
Results obtained are highly encouraging.  In particular the
theory reproduces the trends in the neutralization probabilities of Li, Na,
and K ions that scatter off clean Cu (001) surfaces.  It also agrees
qualitatively with the measured
negative ion fractions of Li and Na ions that interact with low
work function surfaces.  For the case of lithium the theory
predicts the
existence of a peak in the intensity of the optical 2p $\rightarrow$ 2s
transition as a function of the surface work function and this peak
has now been seen in our experiments.

A number of fascinating questions can be posed within this framework.
These questions can be answered by extending the
existing model and its solution to more general situations:

(1) Experiments with other ion species, such as atomic oxygen\refto{Craig},
call for theoretical attention.  The incorporation of additional orbitals and
initial states with different orbital angular momentum into the
Newns-Anderson model and our systematic solution
is straightforward and will not slow down the numerical integrations
significantly.  A theory of Oxygen using the slave-Boson formalism
was presented recently\refto{PN2}.

(2) The incorporation of more realistic target band structure
into the model is also fairly simple.  For example, surface states can be
included as a separate metallic band.  In addition,
experiments on semiconducting targets have been done\refto{Greene}
and these measurements should be reexamined using the many-body theory.

(3) Related to the nature of the band structure are the effects of
the parallel component of the ion velocity\refto{LG,GRLG1}
and the local electronic
structure induced by adsorbates\refto{Kimmel}.
The band structure of the target and the atom-metal matrix elements
$V_{a; k}(z)$ should be recomputed in
the boosted reference frame of the ion.  This effect has already been
studied within the single-particle picture\refto{WZ} and it would be
worthwhile to include these effects in the many-body model.
Local variations in the electrostatic potential due to
the adsorbates also should be included in the calculation to permit more
quantitative comparisons with experiment.
Finally, how do many-body effects change the overlap matrix elements
$V_{a;k}$ between the atom and the target metal?  Can experiments with
clean surfaces further constrain these parameters?

(4) We can include Auger processes in the many-body theory, and now need
to think about how to model these amplitudes in a meaningful way.  One
danger to be avoided is the introduction of more and more parameters
into the theory without a clear understanding of their values.  There
is provocative theoretical\refto{Fonden} work on Auger
processes which we can draw upon to find sensible models for the amplitudes.
In addition, experiments have found secondary electrons resulting from
ion-surface collisions which are consistent with various Auger
processes,\refto{Kempter,BMKKSK,BMKK} but the determination
of absolute cross sections is a difficult experimental problem.
The Auger term complicates the solution of the
many-body equations significantly
because it involves three momenta indices.  Nevertheless,
preliminary work indicates that the numerical problem remains tractable
as long as L, the number of metal states, is not too large.

Questions outside the framework presented here
include:

(5) How does the systematic 1/N solution compare to the
Langreth-Nordlander\refto{LN}
slave-Boson approach?  As both theories start from
the same basic model, the question concerns the limitations of the
approximations made in the 1/N and slave-Boson solutions.  A preliminary
comparison has shown that qualitatively similar behavior is exhibited
by both theories, in that both ``lock out'' additional charge transfer
in the limit of large Coulomb repulsion
as described in the introduction [section (I)].

(6) How do the electron-electron interactions inside the target metal
affect charge-transfer processes?  Is the Landau Fermi liquid approach
adequate?  Finite quasi-particle lifetimes may play a significant role
in the particle-hole sector of the theory.  It
should at least be possible to include phenomenological lifetimes in the
many-body theory.  One class of interesting systems is the heavy
Fermion materials.  The extremely narrow bandwidths (due to large effective
electron masses) should enhance the formation of particle-hole pairs
and strong collective phenomena may occur.
In a more speculative vein, recent work shows that
resonant tunneling in one-dimensional Luttinger
liquids exhibits anomalous behavior\refto{Kane}.  If, as has been suggested
in the case of the high-temperature superconductors, Luttinger liquids
are realized in higher dimensional materials, would any clear
signatures appear in charge transfer experiments?

(7) What are the limitations of the Newns-Anderson model?  In
particular, we know that the atomic orbitals are distorted and
hybridized as the ion approaches the surface.  Do new resonances appear?
Should explicit matrix elements be added to model direct hybridization?
Analysis of experiments on static adsorbed atoms might yield insight into
these questions.

(8) Can the asymptotic formalism\refto{Dorsey} of Dorsey {\it et al.} be
incorporated into the 1/N solution?  This approach might permit more efficient
solutions of the dynamical equations.  It involves a calculation, within
the 1/N approximation, of the static eigenstates of the system at
variable distance $z$.  The dynamical problem can then be solved, for
a series of different perpendicular
velocities, using these states (which need only be
evaluated once).  We already calculate the ground state of the system at
the point of closest approach; perhaps only low-lying states are
needed for an accurate description of the dynamics.

\bigskip
\centerline{Acknowledgements}
\smallskip

We thank A. Dorsey, D. Goodstein, D. Langreth, P. Nordlander, J. Sethna,
and E. Zaremba for helpful discussions.  We also thank P. Nordlander for
providing us with results prior to publication.  This research was supported by
an IBM postdoctoral fellowship (J.B.M), the Cornell Materials Science Center
and the National Science Foundation (grant nos.
NSF-DMR-9022961 and NSF-DMR-9121654). D.R.A. was also supported by the
Swedish Institute and the Sweden-America Foundation.
Additional support for this research was provided by the
Air Force Office of Scientific
Research (grant no. AFOSR-91-0137).

\vfill\eject

%%%%%%%%%%%%%%%%%%%%%%%%%%%%%%%%%%%%%%%%%%%%%%

\head{Figure Captions}

\item{(1)} Schematic of the four sectors kept in the variational many-body
wavefunction.  The Fermi energy is denoted $\epsilon_f$ and the circle
is the hyperthermal atom.
State $|0>$ represents a positive ion with a closed inner shell
and an unperturbed Fermi liquid.  State $|a;k>$, on the
other hand, is a neutral atom with orbital $a$ filled plus a hole at
momentum $k$ in the metal.  It is obtained from the state $|0>$ by the transfer
of an electron from state $k$ in the metal to the atomic orbital $a$ (arrow).
State $|l,k>$ is a particle-hole pair: the
electron has moved from momentum $k$ (creating a hole) to momentum $l$
via a hop to and from the atom.
It is vital to include this next-order term in the calculation because
it plays an important role in the loss-of-memory of the initial incoming state
(see text).
Finally, $|k,q>$ is the state of a negative ion -- the two
valence electrons are in the lowest orbital --
with two holes of momenta $k$ and $q$ left behind in the metal.

\item{(2)} Occupancy of each sector of the $N = 2$ many-body
wavefunction as a function of the atomic position.  Here a lithium atom
interacts with a
metal surface of work function $\Phi = 4.0$ eV.  The band consists of
100 states above and 100 states below the Fermi surface with a
full bandwidth of 8 eV and constant density of states.  The couplings are
determined from the atomic lifetimes calculated in the single-particle
approximation of \Ref{NT} and \Ref{PN} (see section IV).
(a) A positive ion heads inward towards the metal
at $z_f = 20.0$ \AA\ with initial perpendicular velocity
$u_i = 0.04\ au$, bounces at $z_0 = 1.0$ \AA\ , and then departs at a lower
velocity of $u_f = 0.03\ au$.
The final occupancy probabilities are:
$P^+ = 0.2150$ (with and without a particle-hole pair), $P^0 = 0.7838$
(with virtually no excited states),
$P^- = 0.0011$ and the probability for one particle-hole to be formed
during the collision is $0.0439$.
(b) Trajectory leaves the equilibrium ground state at the point of
closest approach with a velocity of $u_f = 0.03 \ au$.
The final occupancy probabilities are:
$P^+ = 0.3453$ (with and without a particle-hole pair), $P^0 = 0.6546$
(also with no excited states),
$P^- = 0.0001$ and the probability for one particle-hole to emerge
from the equilibrium ground state is $0.0024$.

\item{(3)} Measured and predicted neutralization probability $P^0$
of lithium and sodium that scatter off of a clean Cu(001) surface.
The neutral fraction is plotted as a function of perpendicular
velocity.  The scattering geometry is depicted in the inset.

\item{(4)} The variation of single-particle level energies with
distance from the surface.  Shown are the affinity level
and two ionization levels of Li obtained within a simple
single-particle picture (see \Ref{Kimmel2}).  The corresponding
electron affinity and ionization energies of an isolated Li atom
are given on the right hand side of the figure.  Note that $z = 0.0$
\AA\ corresponds to the jellium edge, and the energy zero corresponds
to the vacuum.  The occupied levels of the metal are shown and the Fermi
energy for the clean Cu surface lies $\Phi = 4.59$ eV below the vacuum
energy level.  As the particle approaches the surface,
the single-particle levels broaden into resonances (not shown).

\item{(5)} The measured relative yield of Li(2p)
(solid triangles) versus the work
function shift, $\Delta\Phi$, induced by the deposition of K,
from the clean Cu(001) surface value of 4.59 eV.
Here Li$^+$ is incident on K/Cu(001) with $E_o = 400$ eV
and $\theta _i = 65^\circ$.
Noteworthy features in the data (solid triangles) include
the peak occurring at $\Delta\Phi \approx 1.8$ eV and the large
overall width of the peak.  The solid line represents the prediction
of the many-body model, normalized to the peak value of the measured yield.

\item{(6)} The charge fractions in the scattered flux,
P$^+$ and P$^-$, versus the work function shift, $\Delta\Phi$,
induced by the deposition of Cs.  Here the incident energy of Li$^+$
is $E_o = 400$ eV and the scattering geometry is given by
$\theta _i = 65^\circ, \theta _f = 64^\circ$.  The lines show the
predictions of the many-body model: solid for positive ion fraction, dashed
for negative ion fraction.

\item{(7)} The peak energies $E/E_0$ in the measured energy spectra (open
circles) are plotted as a function of incident energy
$E_0$ and compared to the energies predicted by the classical
trajectory simulation SAFARI (solid triangles and line). Energy loss
through particle-hole pair formation is not included in the
simulations, only energy transfer to the recoiling surface atoms.

\vfill\eject

%%%%%%%%%%%%%%%%%%%%%%%%%%%%%%%%%%%%%%%%%%%%%

\pageno=40
\references

\refis{FD}A. J. Algra, E. V. Loenen, E. P. Th. M. Suurmeijer, and
	A. L. Boers, Radiat. Eff. {\bf 60}, 173 (1982); B. Rasser,
	J. N. M. van Wunnik, and J. Los, Surf Sci. {\bf 118}, 697 (1982).

\refis{Work}P. O. Gartland, Phys. Norv. {\bf 6}, 201 (1972; P. O. Gartland,
	S. Berge, and B. J. Slagsvold, Phys. Norv. {\bf 7}, 39 (1973).

\refis{Shao}Peter Nordlander, Hongxiao Shao, and David C. Langreth,
	unpublished.

\refis{CB}For a review, see D. V. Averin and K. K. Likharev in {\it
	Quantum Effects in Small Disordered Systems}, edited by B. Al'tshuler,
	P. A. Lee, and R. A. Webb (Elsevier, New York, 1990), and
	references therein.

\refis{Chris}C. A. DiRubio, B. H. Cooper, G. A. Kimmel, and R. L.
	McEachern, Nucl. Inst. Meth. Phys. Res. B {\bf 64}, 49
	(1992); C. A. DiRubio, R. L. McEachern, J. G. McLean, and B. H. Cooper,
	in preparation.

\refis{Cyrus}Cyrus Umrigar, private communication.

\refis{JBM}A copy of the C-language program implementing these
	equations, ``nstate.c,'' is available to interested parties.
	Please contact J. B. Marston via e-mail at:
	jbm@yollabolly.physics.brown.edu.

\refis{Yu}M. L. Yu and N. D. Lang, Phys. Rev. Lett. {\bf 50}, 127 (1983).

\refis{Kimmel}G. A. Kimmel, D. M. Goodstein, Z. H. Levine, and
	B. H. Cooper, Phys. Rev. B{\bf 43}, 9403 (1991).

\refis{Geerlings}J. J. C. Geerlings, L. F. Tz. Kwakman, and J. Los,
	Surf. Sci. {\bf 184}, 305 (1987).

\refis{BN1}R. Brako and D. M. Newns, Surf. Sci. {\bf 108}, 253 (1981).

\refis{BN3}R. Brako and D. M. Newns, Rep. Prog. Phys. {\bf 52}, 655 (1989).

\refis{Kondo1}C. M. Varma and Y. Yafet, Phys. Rev. B{\bf 13}, 2950 (1976);
	O. Gunnarsson and K. Schonhammer, Phys. Rev. B{\bf 28},
	4315 (1983).

\refis{WZ}H. Winter and R. Zimny, ``Coherence in Grazing Ion-Surface
	Collisions,'' pp. 283 -- 319 in {\it Coherence in Atomic
	Collision Physics}, Plenum Press, New York, NY 1988; R. Zimny,
	Surf. Sci. {\bf 233}, 333 (1990).

\refis{Dorsey}Alan T. Dorsey, Karsten W. Jacobsen, Zachary H. Levine,
	and John W. Wilkins, Phys. Rev. B{\bf 40}, 3417 (1989).

\refis{Kasai}H. Kasai and A. Okiji, Surf. Sci. {\bf 183}, 147 (1987);
	H. Nakanishi, H. Kasai, and A. Okiji, Surf. Sci. {\bf 197}, 515
	(1988).

\refis{Goldberg}E. C. Goldberg, E. R. Gagliano, and M. C. G. Passeggi,
	Phys. Rev. B{\bf 32}, 4375 (1985).

\refis{Kondo2}N. Andrei, Phys. Rev. Lett. {\bf 45}, 379 (1980);
	N. Andrei and J. Lowenstein, Phys. Rev. Lett. {\bf 46}, 356 (1981);
	P. B. Wiegmann, Zh. Eksp. Teor. Fiz. Pis'ma Red. {\bf 31},
	392 (1980) [JETP Lett. {\bf 31}, 364 (1981)].

\refis{Sebastian}K. L. Sebastian, Phys. Rev. B{\bf 31}, 6976 (1985).

\refis{BN2}R. Brako and D. M. Newns, Solid State Comm. {\bf 55}, 633 (1985).

\refis{SAD}K. W. Sulston, A. T. Amos, and S. G. Davison, Phys. Rev. B{\bf 37},
	9121 (1988); A. T. Amos, K. W. Sulston, and S. G. Davison,
	``Theory of Resonant Charge Transfer in Atom-Surface Scattering'',
	{\it Advances in Chemical Physics, Vol. LXXVI, Molecular Surface
	Interactions}, pp. 335 -- 368, John Wiley \& Sons Ltd. (1989).

\refis{SAD2}K. W. Sulston, A. T. Amos, and S. G. Davison, Surf. Sci.
	{\bf 224}, 543 (1991).

\refis{Nak}Nakanishi {\it et al.}, Surf. Sci. {\bf 242}, 410 (1991);
	Nakanishi {\it et al.}, Surf. Sci. {\bf 216}, 249 (1989).

\refis{LN}David C. Langreth and P. Nordlander, Phys. Rev. B{\bf 43},
	2541 (1991).

\refis{Nordlander}P. Nordlander, private communication.

\refis{PN}P. Nordlander, Phys. Rev. B {\bf 46}, 2584 (1992).

\refis{NT}P. Nordlander and J. C. Tully, Phys. Rev. Lett. {\bf 61}, 990
	(1988); Surf. Sci. {\bf 211/212}, 207 (1989); Phys. Rev. B{\bf 42},
	5564 (1990).

\refis{LG}J. Los and J. J. C. Geerlings, Phys. Reports {\bf 190}, 133 (1990).

\refis{Gauyacq}A. G. Borisov, D. Teillet-Billy and J. P. Gauyacq, unpublished.

\refis{GRLG1}J. J. C. Geerlings, R. Rodink, J. Los, and J. P. Gauyacq,
	Surf. Sci. {\bf 181}, L177 (1987).

\refis{GRLG2}J. J. C. Geerlings, R. Rodink, J. Los, and J. P. Gauyacq,
	Surf. Sci. {\bf 186}, 15 (1987).

\refis{Kempter}H. Schall, H. Brenten, K. H. Knorr, and V. Kempter, Z. Phys.
	D{\bf 16}, 161 (1990).

\refis{Fonden}Tony F\'onden and Andre Zwartkruis, Surf. Sci. {\bf 269/270},
	601 (1992); N. Lorente and R. Monreal, unpublished; R. Zimny,
	Surf. Sci. {\bf 255}, 135 (1991) and unpublished.

\refis{safari}D. M. Goodstein, S. A. Langer, and B. H. Cooper, J. Vac.
	Sci. Technol. A {\bf 6}, 703 (1988).

\refis{Craig}Craig Keller and B. H. Cooper, unpublished.

\refis{Kane}C. Kane and M. Fisher, Phys. Rev. Lett. {\bf 68}, 1220
	(1992) and preprint.

\refis{SHHMK}H. Schall {\it et al.} Surf. Sci. {\bf 210}, 163 (1989).

\refis{BACM}E. R. Behringer, D. R. Andersson, B. H. Cooper, and J. B. Marston,
	to be published.

\refis{BMKKSK}H. Brenten \it et al. \rm Surf. Sci. {\bf 243}, 309 (1991).

\refis{BMKK}H. Brenten \it et al. \rm Nucl. Inst. Meth. {\bf B58}, 328
	(1991).

\refis{Kimmel2}G. A. Kimmel and B. H. Cooper, to be published.

\refis{Ashwin}M. J. Ashwin and D. P. Woodruff, \rm Surf. Sci.
	{\bf 244}, 247 (1991).

\refis{Cooper3}D. L. Adler and B. H. Cooper, Rev. Sci. Instrum. {\bf 59},
	137 (1988).

\refis{Cooper4}R. L. McEachern, D. L. Adler, D. M. Goodstein, G. A. Kimmel,
	B. R. Litt, D. R. Peale, and B. H. Cooper, Rev. Sci. Instrum.
	{\bf 59}, 2560 (1988).

\refis{Cooper5}G. A. Kimmel and B. H. Cooper, to be published.

\refis{PN2}C. C. Hsu, H. Bu, A. Bousetta, J. W. Rabalais, and P. Nordlander,
	Phys. Rev. Lett. {\bf 69}, 188 (1992).

\refis{Greene}Y. Bu, E. F. Greene, and D. K. Stewart, J. Chem. Phys.
	{\bf 92}, 3899 (1990).

\endreferences
\vfill\eject
\end